\documentclass[11pt,USenglish,letterpaper]{article}
\usepackage[T1]{fontenc}
\usepackage[latin9]{inputenc}
\usepackage{amsmath}
\usepackage{todonotes}
\usepackage{hyperref}
\usepackage[affil-it]{authblk}
\usepackage[margin=0.9in]{geometry}
\usepackage[ruled,linesnumbered]{algorithm2e}
\usepackage{babel}
 \usepackage{subcaption}
 \usepackage{float}
\usepackage{mathtools}
\usepackage{numprint}
\usepackage{booktabs}
\usepackage{wrapfig}
\usepackage{graphicx}
\usepackage{placeins}
\usepackage[export]{adjustbox}

\newcommand{\etal}{\textit{et al.}}

\usepackage{fancyhdr}
\newcount\shortyear\newcount\shorthour\newcount\shortminute
\shorthour=\time\divide\shorthour by 60\shortyear=\shorthour
\multiply\shortyear by 60\shortminute=\time\advance\shortminute by
-\shortyear
\shortyear=\year\advance\shortyear by -1900

\def\zeit{\number\shorthour:\ifnum\shortminute<10 0\number\shortminute
\else\number\shortminute\fi}

\usepackage{ifthen}
\newboolean{double_blind}
\setboolean{double_blind}{false}

\begin{document}

\ifthenelse{\boolean{double_blind}}
{
\author{} 
}
{
\author{Patrick Bisenius, Elisabetta Bergamini, Eugenio Angriman, Henning Meyerhenke\thanks{This work is partially supported by German Research Foundation (DFG) grant ME 3619/3-1 within the 
Priority Programme 1736 \emph{Algorithms for Big Data.}}}
}

\title{Computing Top-$k$ Closeness Centrality \\ in Fully-dynamic Graphs}

\date{}

\maketitle

\begin{abstract}
Closeness is a widely-studied centrality measure. Since it requires all pairwise distances, computing closeness for all nodes is infeasible for large real-world networks. However, for many applications, it is only necessary to find the $k$ most central nodes and not all closeness values. Prior work has shown that computing the top-$k$ nodes with highest closeness can be done much faster than computing closeness for all nodes in real-world networks.

However, for networks that evolve over time, no dynamic top-$k$ closeness algorithm exists that improves on static recomputation. In this paper, we present several techniques that allow us to efficiently compute the $k$ nodes with highest (harmonic) closeness after an edge insertion or an edge deletion. Our algorithms use information obtained during earlier computations to omit unnecessary work. However, they do not require asymptotically more memory than the static algorithms (i.\,e., linear in the number of nodes). 

We propose separate algorithms for complex networks (which exhibit the small-world property) and networks with large diameter such as street networks, and we compare them against static recomputation on a variety of real-world networks. On many instances, our dynamic algorithms are two orders of magnitude faster than recomputation; on some large graphs, we even reach average speedups between $10^3$ and $10^4$.

\end{abstract}

\newpage

\section{Introduction}
\label{sec:intro}
\paragraph*{Context and Motivation.}
%
Centrality is a widely used concept in network analysis in order to rank the nodes of a graph by their structural importance.
(Edge centrality measures exist as well, but are not considered here.)
Many centrality definitions include the length of shortest paths, their fraction or their number, the importance
of a node in random walk processes, and many more~\cite[Chap.~7]{Newman2010}.
Different applications may require different centrality measures and none is universal; thus, dozens of
different measures have been spotted in the literature~\cite{DBLP:journals/im/BoldiV14}.

%
In this work, we focus on one of the most widely used measures, \textit{closeness centrality}. 
It is defined as the inverse of the average shortest-path distance. Example applications listed in
previous work~\cite{bergamini2016computing} include facility location, marketing strategies,
and the identification of key infrastructure nodes as well as disease propagation control and crime prevention.

Computing the closeness centrality of a node in an unweighted graph requires a complete breadth-first search (BFS) --- or a complete run of Dijkstra's algorithm for weighted graphs. Moreover, it requires solving the \emph{all-pairs-shortest-path} problem to compute the closeness centrality of each node in the graph. The computational effort for this is often impractical for very large real-world networks.
%
%
For some applications, however, it is enough to compute a list of the $k$ most central nodes. As shown by
Bergamini \etal~\cite{bergamini2016computing}, limiting oneself to this so-called \emph{top-$k$ closeness centrality} 
problem can decrease the computational effort significantly for real-world networks. 
The idea of existing approaches~\cite{borassi2015fast,bergamini2016computing} is to compute upper bounds on the closeness of each node. If we find $k$ nodes whose closeness is higher than the upper bounds on the remaining nodes, we know these nodes have to be the top $k$.

Although these approaches were shown to work very well in practice, they are asymptotically not faster than computing closeness for all nodes. In fact, the most central node in a graph cannot be computed in $\mathcal{O}(|E|^{2 - \epsilon})$ in directed graphs in the worst-case, under reasonable complexity assumptions~\cite{bergamini2016computing}.


Many real-world networks change over time; just think of social networks constantly adding new users (node insertions)
or friendship relations (edge insertions). Terminating a friendship in the social network corresponds to an edge removal.
Similar effects can be seen in the web graph or co-authorship graphs with sliding time windows.
Edge modifications may cause shortest paths to appear or disappear, depending on the modification.
A simple strategy to get the new closeness centralities of each node is then to rerun the static algorithm
on the modified graph, ignoring any information collected by previous runs of the algorithm. 
However, as previous work for related
%
%
\ifthenelse{\boolean{double_blind}}
{
problems~\cite{DBLP:conf/asunam/KasCC13,DBLP:conf/socialcom/GreenMB12,DBLP:journals/im/BergaminiM16} 
}
{
problems~\cite{DBLP:conf/asunam/KasCC13,DBLP:journals/im/BergaminiM16,DBLP:conf/wea/BergaminiMOS17,DBLP:conf/socialcom/GreenMB12} 
}
has shown, it is usually much more efficient 
in practice to make more localized updates to the centrality values.

\paragraph*{Outline and Contribution.}
This paper presents dynamic algorithms for top-$k$ closeness that handle both edge insertions and edge removals,
see Section~\ref{sec:algorithm}. Our new algorithms are based on the static algorithms proposed by Bergamini \etal~\cite{DBLP:journals/corr/BergaminiBCMM17} (and in~\cite{bergamini2016computing}, in a preliminary version). Since traditional closeness does not apply to disconnected graphs, our algorithms compute a variant called \textit{harmonic closeness}, which has been shown in~\cite{DBLP:journals/im/BoldiV14} to satisfy all axioms presented in the same paper.
However, our algorithms can be easily adapted to Bavelas's definition of closeness as well.

Our algorithms reuse information obtained by an initial run of the static algorithm and try to skip the recomputation of closeness centralities for nodes that are unaffected by modifications of the graph. In contrast to other dynamic algorithms for closeness centrality, it is not required to compute the exact closeness centralities of all nodes in the initial graph, making it possible to target networks with hundreds of millions of edges. Moreover, we specifically design our algorithms to use only a linear amount of additional memory, since a quadratic memory footprint (typical of most existing dynamic algorithms for problems based on shortest paths) would be impractical for large instances.
In experiments we obtain significant speedups compared to static recomputation. For example, for $k=10$, our average speedup (geometric mean over the tested instances) is about \numprint{76} for insertions in undirected complex networks. For deletions in directed street networks, we reach an average speedup of \numprint{743}. Also, our experiments show some non-trivial results: deletions are mostly faster than insertions and speedups increase with $k$ for complex networks, whereas they decrease as $k$ increases in street networks.

\section{Preliminaries}
\subsection{Notation and Problem Definition}
Let $G$ be an unweighted graph (either directed or undirected) with $n$ nodes and $m$ edges. We use $d(u, v)$ to denote the shortest-path distance between two nodes $u$ and $v$. The set of nodes at distance $i$ from $u$ is denoted by $N_i(u) := \{ v : d(u, v) = i\}$ and its cardinality by $n_i(u)$. The reachable nodes $R(u) := \{ v : d(u, v) < + \infty \}$ are the nodes with finite distance from $u$ (we denote their cardinality by $r(u)$).
According to Bavelas's definition, the closeness centrality of node $u$ in a (strongly) connected graph is $\frac{n-1}{\sum_{v \in V} d(u, v)}$~\cite{Bavelas1950}. For disconnected graphs, this quantity is not defined, since the denominator would be infinity for all nodes. We consider a variant of closeness called \textit{harmonic closeness centrality}~\cite{DBLP:journals/im/BoldiV14}, defined as follows:
$$c(u) := \sum_{v \in V, \ v \neq u} \frac{1}{d(u, v)} \ .$$
In addition to extending to disconnected graphs in a very natural way, harmonic closeness has been shown in~\cite{DBLP:journals/im/BoldiV14} to satisfy all axioms presented in the same paper (i.e., size, density and score monotonicity). 

When talking about dynamic graphs, we refer to $G$ as the graph before the edge update, and to $G'$ as the modified graph. Similarly, $d$ is the distance on $G$ and $d'$ the distance on $G'$.

\subsection{Related Work}
\label{sec:rel-work}
\paragraph{Static Algorithms.}
Closeness centrality is based on pairwise shortest-path distances. 
In unweighted graphs, these are usually computed by running a BFS (Breadth-First Search) from each node, requiring $\Theta(nm)$ time.
Unfortunately, this is impractical for large networks with hundreds of millions of nodes and edges. For this reason, several approximation algorithms have been proposed over the years. The simple approach by Eppstein and Wang samples a set of source nodes, runs a BFS from them and uses the computed distances to extrapolate the closeness of the other nodes~\cite{DBLP:journals/jgaa/EppsteinW04}. If the graph has a bounded diameter, this approach delivers an additive error guarantee with high probability.
Cohen et al.~\cite{DBLP:conf/cosn/CohenDPW14} improve on this approach by combining the sampling with a new $3$-approximation algorithm. This provides an estimate $\tilde c(v)$ of the centrality of each node $v$ such that $P\left(\left|\frac{1}{\tilde c(v)}-\frac{1}{c(v)}\right| \geq \frac{\epsilon}{c(v)} \right) \leq 2e^{-\Omega\left(l\epsilon^3\right)}$, where $l$ is the number of samples.
The recent result by Chechik et al.~\cite{DBLP:conf/approx/ChechikCK15} allows to approximate closeness centrality with a coefficient of variation of $\epsilon$ using $O(\epsilon^{-2})$ BFS computations. Alternatively, one can make the probability that the maximum relative error exceeds $\epsilon$ polynomially small by using $O(\epsilon^{-2} \log n)$ BFS computations.
Although approximation algorithms can often provide scores that are close to the real ones, they may fail at preserving the ranking, in particular for nodes with similar closeness. For example, it was argued in~\cite{DBLP:journals/corr/BergaminiBCMM17} that the algorithm by Chechik et al.~\cite{DBLP:conf/approx/ChechikCK15} would require $O(m n^2)$ time to deliver a reliable ranking in a small-world network.

For this reason, the problem of preserving the ranking of the top-$k$ nodes with highest closeness has been considered, both exactly~\cite{DBLP:conf/icde/OlsenLH14}, with high probability~\cite{DBLP:conf/faw/OkamotoCL08} and through heuristics~\cite{DBLP:journals/ipl/MerrerST14,Lim11onlineestimating}. 

In particular, the exact approaches proposed in~\cite{DBLP:journals/corr/BergaminiBCMM17} (and in~\cite{bergamini2016computing}, in a preliminary version) have been shown to outperform existing algorithms for finding the top-$k$ nodes with highest closeness centrality. More specifically, the authors of this prior work propose a method (\textsf{NBCut}) that works best for networks with small diameter (such as complex networks), and a method (\textsf{NBBound}) that is more efficient on networks with relatively large diameter (such as street networks). Since our new dynamic algorithms build on these two approaches, we explain them in more detail in Section~\ref{sec:top-k-static}. Notice that these are \textit{exact} approaches, i.e., they find the $k$ nodes with highest closeness and their exact closeness values. Similarly, also our dynamic algorithms are exact.

\paragraph{Dynamic Algorithms.}
Many of today's networks continuously change over time. Recomputing the centrality values after each edge modification may be too expensive for large instances.
Dynamic algorithms try to update some properties of the graph by limiting the computations to a subset of the nodes and edges. For updating the closeness of all nodes, a simple algorithm has been proposed by Kas et al.~\cite{DBLP:conf/asunam/KasCC13}. The authors use a dynamic algorithm by Ramalingam and Reps~\cite{DBLP:journals/tcs/RamalingamR96} for updating pairwise distances and either increase or decrease the closeness of nodes whose distance has changed. A major limitation of this approach is its memory requirement of $\Theta(n^2)$, which is too expensive for large networks.

For unweighted graphs, Sariy{\"{u}}ce et al.~\cite{DBLP:conf/bigdataconf/SariyuceKSC13} present optimizations that make the dynamic algorithm more efficient on complex networks. In particular, they show that the recomputation of closeness can be skipped for nodes $s$ such that $|d(s, u) - d(s, v)| = 1$ (where $u$ and $v$ are the endpoints of the newly inserted or deleted edge). Also, they divide the graph into biconnected components and show that nodes outside the biconnected component of $(u, v)$ can also be skipped. Finally, they notice that nodes with the same neighborhood have the same closeness, and therefore (re)computing it for only one of the nodes is sufficient.
Differently from the algorithm by Kas et al.~\cite{DBLP:conf/asunam/KasCC13}, the one by Sariy{\"{u}}ce et al.~\cite{sariyuce2013incremental} does not store pairwise distances, resulting in a memory requirement of $\Theta(n)$. Nevertheless, both algorithms require to compute exact closeness centrality at least once on the initial graph, which might be impractical.

\subsection{Static Top-$k$ Closeness}
\label{sec:top-k-static}
The textbook algorithm would compute closeness by running a BFS from each node. 
Both top-$k$ closeness algorithms proposed in~\cite{DBLP:journals/corr/BergaminiBCMM17} try to reduce the running time of the textbook algorithm by exploiting upper bounds on the closeness. More precisely, \textsf{NBCut} reduces the work done during the BFSs, but not their number (i.\,e., it starts a BFS from each node). \textsf{NBBound}, on the contrary, runs complete BFSs, but on a limited number of nodes. Although both algorithms were proposed using Bavelas's definition of closeness, it was shown in~\cite{DBLP:journals/corr/BergaminiBCMM17} how they can be adapted to harmonic closeness. In the following, we describe them based on this adaptation.

\begin{figure}[tb]
\centering
\includegraphics[width = 0.52\textwidth]{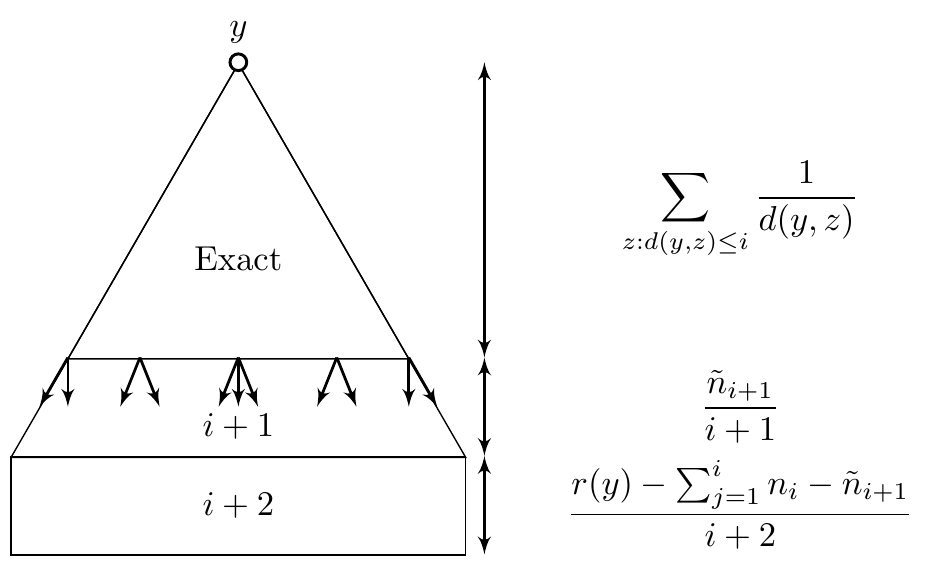}
\caption{Upper bound on $c(y)$ computed by the \textsf{NBCut} algorithm. For nodes up to distance $d_{cut}(y)$ we know the exact distance. Then, $\tilde{n}_{d_{cut}(y) + 1}$ nodes are assumed to be at distance $d_{cut}(y) + 1$ and the remaining at distance $d_{cut}(y) + 2$.}
\vskip -6pt
\label{fig:bfscut}
\end{figure}
\paragraph{\textsf{NBCut} Algorithm for Complex Networks.}
Assume we are performing a BFS from a node $y$, and we have visited all nodes up to distance $i$. We know that all remaining nodes have at least distance $i+1$, otherwise they would have been visited already. Also, we know that at most $\tilde{n}_{i+1} := \sum_{w \in N_i(y)} \mathsf{degree}(w)$ can be at distance $i + 1$, since all nodes at distance $i+1$ must have at least one neighbor at distance $i$ from $y$. Thus, all remaining nodes have to be at distance at least $i + 2$. Based on this observation, we can define the following upper bound on $c(y)$: 
\begin{equation}
\label{eq:bfscut}
\tilde{c}(y) := \sum_{z : d(y, z) \leq i} \frac{1}{d(y, z)} + \frac{\tilde{n}_{i + 1}}{i+1} + \frac{r(y) - \sum_{j=1}^{i} n_{j} - \tilde{n}_{i + 1}}{i+2},
\end{equation}
where the first term in the sum accounts for the nodes for which we computed the exact distance, the second term assumes that exactly $\tilde{n}_{i+1}$ nodes are at distance $i+1$, and the third term assumes that all remaining nodes are at distance $d+2$ (see Figure~\ref{fig:bfscut}). Notice that the more nodes are visited during the BFS, the tighter the bound is (if all nodes are visited, $\tilde{c}(y) = c(y)$).  Thus, the \textsf{NBCut} algorithm (Algorithm~\ref{algo:main2} in Appendix~\ref{app:pseudocodes}) works as follows: assume that we already computed the \textit{exact} closeness for at least $k$ nodes, and let $x_k$ be the $k$-th highest closeness value found so far. While running a BFS from node $y$, we compute the bound of Eq.~(\ref{eq:bfscut}). If at some point $\tilde{c}(y) < x_k$, we can interrupt the search from $y$, since $y$ cannot be one of the top-$k$ nodes. We call the distance $i$ at which we interrupt the BFS the \textit{cutoff distance}, and we refer to it as $d_{cut}(y)$. Also, we refer to this extended BFS that computes $\tilde{c}(y)$ and returns when $\tilde{c}(y) < x_k$ as \textsf{BFScut} (Line~\ref{line:bfscut}). If the exact closeness $c(y)$ is actually larger than $x_k$, we run a complete BFS and replace the current $k$-th node with highest closeness with $y$ (and set \textsf{isExact}$(y)$ to true in Line~\ref{line:bfscut}). 
Clearly, the order in which the nodes are processed is very important: ideally we would like to process them in order of decreasing closeness. In~\cite{DBLP:journals/corr/BergaminiBCMM17}, the authors show that a measure based on the number of walks (the ordering $O$ in Algorithm~\ref{algo:main2}) works quite well. 

Notice that Eq.~(\ref{eq:bfscut}) requires the number $r(y)$ of nodes reachable from $y$. In undirected graphs, this is the size of the connected component of $y$, which can be easily computed in a preprocessing step. In directed graphs, the authors of~\cite{DBLP:journals/corr/BergaminiBCMM17} propose an upper bound based on a topological sorting of the strongly-connected-components DAG (SCC DAG).

\paragraph{\textsf{NBBound} Algorithm for Street Networks.}
\textsf{NBBound} initially computes an upper bound $\tilde{c}$ on the closeness of each node and enters the nodes into a priority queue $Q$, sorted by their value of $\tilde{c}$ (Line~\ref{line:lbounds} of Algorithm~\ref{alg:main} in Appendix~\ref{app:pseudocodes}). Notice that this is unrelated to the $\tilde{c}$ computed by \textsf{BFSCut} (see~\cite{DBLP:journals/corr/BergaminiBCMM17} for more details). Then (Lines~\ref{line:while-start}-\ref{line:while-end}), the node $v$ with highest $\tilde{c}$ is extracted and its actual closeness $c(v)$ is computed using the \texttt{BFSbound} function. In addition to computing the exact closeness of $v$, this function also modifies the upper bounds $\tilde{c}$ of the other nodes.
In particular, let $w$ and $y$ be any two nodes visited during the BFS from $v$. Then, it can be proven~\cite{DBLP:journals/corr/BergaminiBCMM17} that $d(w, y) \geq |d(v, w) - d(v, y)|$. By assuming that $d(w, y) = |d(v, w) - d(v, y)|,\ \forall w,y \in V$, we get an upper bound on the closeness of all nodes. If, for some node $w$, this upper bound is tighter than the current $\tilde{c}$, then $\tilde{c}$ and the priority of $w$ in $Q$ are updated.
The algorithm terminates when $k$ nodes are found such that their exact closeness is higher than the upper bound on the closeness of the other nodes (Line~\ref{line:return}).

\section{Dynamic Top-$k$ Closeness Centrality}
\label{sec:algorithm}
Let us assume an edge $(u,v)$ has been inserted into or deleted from the graph. Our goal is to update the list of the top-$k$ nodes and their closeness values faster than computing it from scratch with \textsf{NBCut} or \textsf{NBBound}. In the following, we first show how we update the number of reachable nodes (Section~\ref{sec:reach}) and compute the set of affected nodes (Section~\ref{sec:affected}), which is a preprocessing necessary for both edge insertions and deletions. Then, we consider edge insertions and deletions separately in Section~\ref{sec:insertions} and Section~\ref{sec:deletions}. 

\subsection{Updating the Number of Reachable Nodes}
\label{sec:reach}
The upper bound in Eq.~(\ref{eq:bfscut}) requires the number of reachable nodes $r(u)$ (or an upper bound on $r(u)$). For undirected graphs, this is simply the number of nodes in the connected component of $u$. Instead of recomputing connected components from scratch after each update, we use a simple dynamic algorithm similar to the one presented in~\cite{EdigerRBM13computational}.
 Initially, we build a spanning forest of $G$. When an edge $(u,v)$ is inserted, we check whether $u$ and $v$ belong to the same component and if not, we merge the components and $(u,v)$ becomes part of the forest. 
 If we delete an edge $(u, v)$ that is part of the forest, we then run a pruned BFS from $u$ and interrupt it as soon as we hit $v$ (and, if we do not hit $v$, we split the components).
 If, in turn, we delete an edge $(u,v)$ that is not part of the spanning forest, we know there has to be another path between $u$ and $v$ and we are done.

For directed graphs, we mentioned in Section~\ref{sec:top-k-static} that a bound based on a topological sorting of the SCC DAG was proposed in~\cite{DBLP:journals/corr/BergaminiBCMM17}. Since preliminary experiments showed that recomputing this after each update was a bottleneck for the dynamic algorithm, we replace this bound with the number of nodes in the weakly-connected component. This does not affect the correctness of the algorithm and can be updated much more efficiently using basically the same algorithm we use for updating the connected components in undirected graphs (we simply need to ignore the direction of the edges).  

\subsection{Finding Affected Nodes}
\label{sec:affected}
When an edge $(u, v)$ is inserted or deleted from $G$, some nodes might increase or decrease their closeness centrality. We call such nodes \textit{affected}. More precisely, the set $A$ of affected nodes is defined as $A := \{y \in V : \exists w \in V \text{ such that } d'(y, w) \neq d(y, w) \}$. It is easy to see that, if $d'(y, w) \neq d(y, w)$, then either $d'(y, u) \neq  d(y, u)$ or $d'(y, v) \neq d(y, v)$. Indeed, if the distances from $y$ to both $u$ and $v$ stay the same, it means that the insertion or deletion of $(u, v)$ does not move any node up or down the BFS DAG rooted in $y$.
Thus, the set of affected nodes can be easily identified by running two BFSs from $u$ (one on $G$ and one on $G'$) and two BFSs from $v$ (one on $G$ and one on $G'$). If the graph is directed, the BFSs have to be run on $G$ and $G'$ transposed, since we are interested in the nodes that change their distance \textit{to} $u$ or $v$. Once we know all the distances to $u$ and $v$ in $G$ and $G'$, we can simply compare them and mark all nodes $y$ such that $d'(y, u) \neq  d(y, u)$ or $d'(y, v) \neq d(y, v)$ as affected.

We know that the closeness of all unaffected nodes $v$ does not change. Therefore, the previously computed upper bound $\tilde{c}$ and, if it was computed, the previous exact closeness value $c(v)$ are still valid.

\subsection{Update after an edge insertion}
\label{sec:insertions}
We first focus on updating the \textsf{NBCut} top-$k$ closeness algorithm.
Before the insertion, we assume that, for each node $y$, we know $\tilde{c}(y)$,  $d_{cut}(y)$ and \textsf{isExact}$(y)$ computed by the function \textsf{BFScut}. Also, we assume the nodes are sorted by their $\tilde{c}$ value in a priority queue $Q$.
After an insertion, affected nodes increase their closeness. One first simple strategy would be to run \textsf{BFScut} from each affected node, using as $x_k$ the current node with the $k$-th highest closeness. This would already save some time compared to the static algorithm, since no work is performed for unaffected nodes. In the following, we propose some further improvements. 
Algorithm~\ref{algo:insertions} in Appendix~\ref{app:pseudocodes} shows the pseudocode of the dynamic algorithm.
Initially, Line~\ref{line:preproc1} and Line~\ref{line:preproc2} update the number of reachable nodes and compute the set $A$ of affected nodes, as described in Section~\ref{sec:reach} and Section~\ref{sec:affected}, respectively.
Then (Lines \ref{line:removeStart}-\ref{line:removeEnd}), the affected nodes that are among the top $k$ are removed from \textsf{TopK}. For each affected node $y$, Lines~\ref{line:opt1}-\ref{line:opt2} try to efficiently update $\tilde{c}$, in order to avoid a new \textsf{BFScut} (as we will see in the following). If this is not possible, Lines~\ref{line:dynbfscut1}-\ref{line:dynbfscut2} run \textsf{BFScut} and update \textsf{TopK} as in the static algorithm.

\paragraph{Skipping Far-away Nodes.}
Let $y$ be an affected node such that $\textsf{isExact}(y)$ is false, i.e., a node for which \textsf{BFScut} has been interrupted at some cutoff level $d_{cut}(y)$. W.l.o.g., let us assume that $d(y, u) < d(y, v)$ if $G$ is undirected. We recall that we know $d(\cdot, u)$ from the identification of the affected nodes described in Section~\ref{sec:affected}.

\begin{figure}[tb]
\begin{center}
\includegraphics[width = 0.36\textwidth]{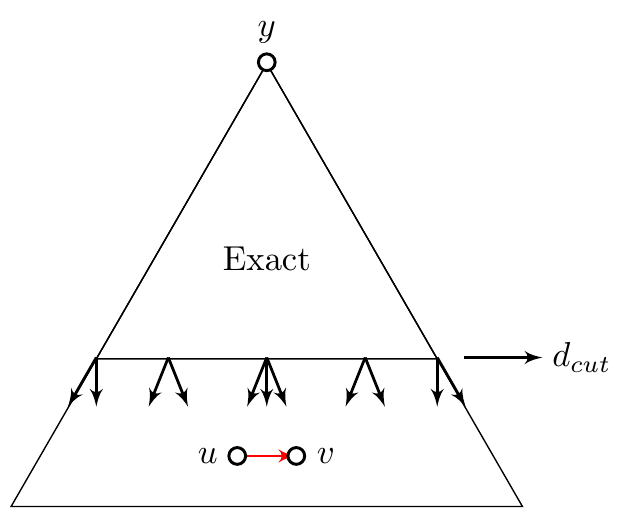}
\hspace{1.5cm}
\includegraphics[width = 0.36\textwidth]{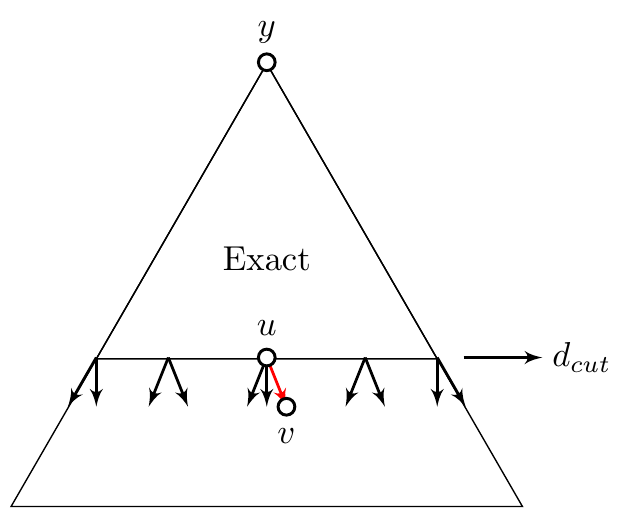}
\caption{Left: $u$ and $v$ are far-away nodes for $y$. Right: $u$ is a boundary node.}
\label{fig:optimizations}
\end{center}
\vspace{-4ex}
\end{figure}

If $d(y, u) > d_{cut}(y)$ and $r'(y) = r(y)$, the upper bound $\tilde{c}(y)$ is still valid. Figure~\ref{fig:optimizations} (left) shows this case. The reason for this is that $u$ has not been visited by \textsf{BFScut} and therefore the existence of edge $(u,v)$ does not affect the bound in Eq.~(\ref{eq:bfscut}). If $d(y, u) > d_{cut}(y)$ but $r'(y) \neq r(y)$ (i.e., the insertion has increased the number of nodes reachable from $y$), we can simply replace $r(y)$ with $r'(y)$ in Eq.~(\ref{eq:bfscut}), obtaining $\tilde{c}'(y) = \tilde{c}(y) - \frac{r(y)}{d_{cut}(y)+2} + \frac{r'(y)}{d_{cut}(y)+2}$.
If $\tilde{c}(y) < x_k$, $y$ can therefore be skipped and no \textsf{BFScut} on $G'$ has to be run from it. We call such nodes \textit{far-away nodes} (Line~\ref{line:faraway} of Algorithm~\ref{algo:insertions} in Appendix~\ref{app:pseudocodes}).

\paragraph{Skipping Boundary Nodes.}
If $d(y, u)$ is equal to $d_{cut}(y)$ (Figure~\ref{fig:optimizations}, right), the bound in Eq.~(\ref{eq:bfscut}) is affected, as the degree of $u$ changes (we recall that $\tilde{n}_{i+1} := \sum_{w \in N_i(y)} \mathsf{degree}(w)$). In particular, $\tilde{n}'_{d_{cut}(y) + 1}$ after the insertion is equal to $\tilde{n}_{d_{cut}(y) + 1} + 1$, since the degree of $u$ has increased by one.
Thus, we can easily compute the new bound from the old one without running a new \textsf{BFScut} from $y$ as follows (we use $d_{cut}$ instead of $d_{cut}(y)$ for simplicity):
\[
\begin{split}
\tilde{c}'(y) - \tilde{c}(y) & =   \frac{\tilde{n}_{d_{cut} + 1} + 1}{d_{cut}+1} + \frac{r'(u) - \sum_{j=1}^{d_{cut}} n_{j} - \tilde{n}_{d_{cut} + 1} - 1}{d_{cut}+2} - \frac{\tilde{n}_{d_{cut} + 1}}{d_{cut}+1} - \frac{r(u) - \sum_{j=1}^{d_{cut}} n_{j} - \tilde{n}_{d_{cut} + 1}}{d_{cut}+2}\\
 & = \frac{1}{d_{cut} + 1} - \frac{r(y) - r'(y) + 1}{d_{cut} + 2}\ .
\end{split}
\]
Boundary nodes are handled in Line~\ref{line:boundary} of Algorithm~\ref{algo:insertions} in Appendix~\ref{app:pseudocodes}.
\paragraph{Distance-based Bounds.}
The improvements described in the previous two paragraphs do not apply to affected nodes $y$ for which $d(y, u) < d_{cut}(y)$.
Let $z$ be any node such that $d'(y, z) < d(y, z)$ (if $y$ is affected, there has to exist such a node). Since all new shortest paths have to go through $(u, v)$ (and thus through $u$), we can write $d'(y,z)$ as $d'(y, u) + d'(u, z) = d(y, u) + d'(u, z)$, since the distance from $y$ to $u$ cannot change as a consequence of the insertion of $(u, v)$. As for $d(y, z)$, there are two options: either $u$ was part of a shortest path from $y$ to $z$ also before the insertion -- and thus $d(y, z) = d(y, u) + d(u, z)$, or there was a shorter path from $y$ to $z$ not going through $u$ -- and therefore $d(y, z) < d(y, u) + d(u, z)$. In both cases, we can say that $d(y, z) \leq d(y, u) + d(u, z)$. Putting this together, we get 
$
\frac{1}{d'(y, z)} - \frac{1}{d(y,z)} \leq \frac{1}{d(y, u) + d'(u, z)} - \frac{1}{d(y, u) + d(u, z)}
$. 
Thus:

\begin{equation}
\label{eq:ubound}
\begin{split}
c'(y) - c(y) & =  \sum_{z \in V} \left( \frac{1}{d'(y, z)} -  \frac{1}{d(y, z)} \right)\\
 &   \leq \sum_{z \in V} \left(  \frac{1}{d(y, u) + d'(u, z)} - \frac{1}{d(y, u) + d(u, z)} \right)  \\
 & \leq \sum_{i = 1}^{\mathsf{diam}}  \frac{1}{i + d(y, u)}  \left( {n}'_i(u) -n_i(u) \right)
\end{split}
\end{equation}
where \textsf{diam} is the diameter of $G$. Notice that Eq.~(\ref{eq:ubound}) implies that, if $\tilde{c}(y)$ is an upper bound on $c(y)$, then $\tilde{c}'(y) := \tilde{c}(y) +  \sum_{i = 1}^{\mathsf{diam}} \frac{1}{i + d(y, u)}  \left( {n}'_i(u) -n_i(u) \right)$ is an upper bound on $c'(y)$. The values $\left( {n}'_i(u) -n_i(u) \right)$ can be easily computed  with one BFS from $u$ in $G$ and $G'$ (this can also be combined with the BFSs we run to identify the affected node, see Section~\ref{sec:affected}). Then, for each affected node that is neither a boundary node nor a far-away node, we compute a new upper bound as in Eq.~(\ref{eq:ubound}). If this is still smaller than $x_k$, no \textsf{BFScut} has to be performed from the node. Notice that the computation of the new bound requires $\Theta(\textsf{diam})$ operations. Since the diameter in complex networks is very small (often assumed to be constant), this is much faster than running a \textsf{BFScut}, which can take up to $\Theta(n + m)$ time. The distance-based bounds are computed in Line~\ref{line:dbbound1} of Algorithm~\ref{algo:insertions}. In Line~\ref{line:dbbound2}, we set $d_{cut}(y)$ to \textsf{diam} to indicate that the current bound is not a result of a \textsf{BFScut} and should not be used in the future to skip far-away or boundary nodes. 

\paragraph{Updating \textsf{NBBound}.}
So far we described how to update top-$k$ closeness assuming that \textsf{NBCut} has been run on the initial graph. We recall that \textsf{NBBound} only runs complete BFSs, until we find $k$ nodes whose closeness is higher than the upper bounds on the remaining nodes. Thus, there is no cutoff threshold that we can use to skip far-away or boundary nodes. However, we can still make some considerations. First, also in this case $c(y)$ and $\tilde{c}(y)$ of unaffected nodes are still valid and do not need to be changed. Also, the distance-based bounds described in the previous paragraph can be applied to \textsf{NBBound} as well.
If there are nodes $y$ whose new bound is higher than the $k$-th highest closeness value, we run a BFS from $y$. We stop when there is no affected node left whose $\tilde{c}$ is higher than $\mathsf{TopK}[k]$, similarly to the static algorithm. Algorithm~\ref{algo:insertions2} in Appendix~\ref{app:pseudocodes} shows the pseudocode.
\subsection{Update After an Edge Deletion}
\label{sec:deletions}
In some sense, edge deletions are easier to handle than edge insertions. Indeed, since closeness can only decrease as a consequence of a deletion, nothing needs to be done in case none of the top-$k$ nodes is affected (whereas for insertions there could be nodes that increase their closeness and ``overtake'' the previous top $k$). Also, notice that for affected nodes the previous upper bounds are still valid (although they might be less tight). If for some affected node $y$ we know the exact closeness $c(y)$ before the insertion, this becomes now an upper bound on the closeness of $y$ in the new graph (thus, we set $\mathsf{isExact}(y)$ to true).

Once this is done, the algorithm goes through the nodes in $Q$ ordered by $\tilde{c}$. If we find some node $y$ for which $\mathsf{isExact}(y) = \text{false}$, then we compute its new $c'(y)$ and update its priority in $Q$. We stop when $\mathsf{isExact}$ is true for the first $k$ nodes in $Q$. Notice that this approach works for both \textsf{NBCut} and \textsf{NBBound}. Algorithm~\ref{algo:deletions} in Appendix~\ref{app:pseudocodes} shows the pseudocode for deletions, based in \textsf{NBCut}. The one based on \textsf{NBBound} is basically the same, with the difference that in Line~\ref{line:bfscut-del} we do not run a \textsf{BFScut}, but a \textsf{BFSbound}, as in Algorithm~\ref{alg:main}.

\subsection{Running Times and Memory Requirements}
\label{sec:time}
The update of the number of reachable nodes described in Section~\ref{sec:reach} takes $O(n + m)$ in the worst case -- the time to run a BFS from scratch. Computing the set of affected nodes takes $\Theta(n + m)$ time, since we need to run a BFS from $u$ in $G$ and $G'$ (in the graph is undirected, also from $v$).
Then, the algorithms described in Section~\ref{sec:insertions} and Section~\ref{sec:deletions} have to run, in the worst case, a \textsf{BFScut} (or \textsf{BFSbound}, if we are considering the algorithm based on \textsf{NBBound}) for each affected node. Since the worst-case running time of both \textsf{BFScut} and \textsf{BFSbound} is $O(n + m)$~\cite{DBLP:journals/corr/BergaminiBCMM17}, in the worst case the running time of the dynamic algorithms is $O(|A|(n + m))$, where $|A|$ is the number of affected nodes. However, we will see in Section~\ref{sec:experiments} that the number of calls to \textsf{BFScut} is usually a small fraction of the total number of affected nodes (which is, in turn, often only a small fraction of the total number of nodes). 

Concerning memory, our algorithms need to store the bound $\tilde{c}(y)$, \textsf{isExact}$(y)$, the number $r(y)$ of reachable nodes and the cutoff distance $d_{cut}(y)$, for each $v \in V$ (as well as the list \textsf{TopK} with the $k$ nodes with maximum closeness). This requires only $\Theta(n)$ memory, which is asymptotically the same as the static top-$k$ algorithms.

\section{Experiments}
\label{sec:experiments}
\subsection{Experimental Setup}
\setlength{\tabcolsep}{4pt}
\paragraph*{Data sets.} We test our algorithms on numerous directed and undirected real-world complex networks (e.\,g., social networks or web graphs) and street networks. 
All of them can be retrieved from the public repositories SNAP (\url{snap.stanford.edu}), LASAGNE (\url{lasagne-unifi.sourceforge.net}), KONECT (\url{konect.uni-koblenz.de/networks}) and GEOFABRIC (\url{download.geofabrik.de}); the graphs are listed in Tables~\ref{tbl:datasetDirected},~\ref{tbl:datasetUndirected} and~\ref{tbl:streetNetworks} in Appendix~\ref{app:tables}. Since the street networks in Table~\ref{tbl:streetNetworks} are all directed, we also consider an undirected version of them, by ignoring the direction of edges. For each tested graph, we either add or remove 100 edges one at a time and run the dynamic algorithm after each update. Due to time constraints, we only run the static algorithm once every 10 updates (this does not affect results considerably, since the running time of the static algorithm is always approximately the same). For edge insertions we remove 100 random edges from the original graph before running the algorithms, and then add them one-by-one, whereas for removals we just delete 100 random edges.

\paragraph*{Implementation and settings.} Since \textsf{NBCut} outperforms \textsf{NBBound} on complex networks, we only use \textsf{NBCut} and the new dynamic algorithm based on it (Algorithms~\ref{algo:insertions} and~\ref{algo:deletions} in Appendix~\ref{app:pseudocodes}) for our experiments on complex networks. Similarly, we only use \textsf{NBBound} and the dynamic algorithm based on it for our experiments on street networks (Algorithm~\ref{alg:main} in Appendix~\ref{app:pseudocodes}). We recall that our dynamic algorithm for directed graphs uses the number of nodes in the weakly connected components instead of the bound originally proposed in~\cite{DBLP:journals/corr/BergaminiBCMM17}. However, for the static case, we use the original bound, in order to make a fair comparison with previous work. We recall that all algorithms are \textit{exact}, i.e., they find the $k$ nodes with highest closeness and their exact scores, so they only differ in their running time and not in the results they find.

The machine we used for our experiments is a shared-memory server equipped with 256 GB RAM and 2 x 8 Intel(R) Xeon(R) E5-2680 cores at 2.7 GHz of which we use only one since we executed our algorithms sequentially (i.\,e., using a single thread).
The code has been written in C++ and uses the open-source \textit{NetworKit} framework~\cite{Staudt2014}. We plan to publish our code in future releases of the package.


\subsection{Speedups on Recomputation}

\paragraph{Dynamic Complex Networks.}
First, we study the effect of the optimizations proposed in Section~\ref{sec:insertions}. Table~\ref{tbl:optimizationImpact} in Appendix~\ref{app:optimizations} contains the average number of affected nodes over the tested edge insertions and the percentage of nodes skipped due to each of the optimizations in undirected graphs (results for $k=10$). 
The average number of affected nodes is never higher than $34\%$ of the total number of nodes. Also, in all graphs, skipping far-away nodes allows us to ignore the vast majority of affected nodes after an update.
Combined with the cheap updates for boundary nodes and applying the distance-based bounds, we need to run a new \textsf{BFScut} for less than 1\% of the affected nodes on most graphs. 
Notice that the column \textsf{BFScuts} contains the percentage of \textit{affected nodes} for which we run a new \textsf{BFScut}. The percentage of the total number of nodes is therefore much smaller.
Table~\ref{tbl:optimizationImpactDirected} in Appendix~\ref{app:optimizations} shows the results for the directed case. 
Compared to undirected graphs, insertions typically affect smaller portions of the graph (average values at most $\approx 7\%$). Among them, usually a smaller percentage (compared to undirected graphs) are far-away nodes, probably because mostly nodes that are very close to the inserted edge are affected. However, several nodes are skipped because of the distance-based bounds, resulting in a very small number of BFSs. The highest numbers are for \texttt{p2p-Gnutella08}, for which the \textsf{BFScuts} are $\approx 31\%$ of the affected nodes and $\approx 1.5 \%$ of the total.

The speedups (ratio between the running times) of our dynamic algorithm on the static one for complex networks are summarized in the lower part of Figure~\ref{fig:summary}.
Table~\ref{tbl:speedupsInsertionDirected} and Table~\ref{tbl:speedupsInsertionUndirected} in Appendix~\ref{app:complex} report the detailed values for insertions in directed and undirected graphs, respectively. 
For undirected networks, the geometric mean of the speedups (over the 100 edge insertions) are always at least in the double-digit range (with the only exception of \texttt{Mus\_musculus}, where the average speedup for $k=1$ is $8.4$). 
Also, the speedups grow for bigger values of $k$, reaching an average speedup (over all tested undirected networks) of $123$ for $k = 100$. Although for directed graphs the average speedups vary a lot (from $\approx 10$ for \texttt{as-caida20071105} to $\approx 3368$ for \texttt{web-Stanford}, for $k=1$), the results are mostly even better than for the undirected case: the average speedups over all tested networks are $62$ for $k=1$, $93$ for $k=10$ and $174$ for $k=100$. 
This can be explained by the smaller number of affected nodes in directed networks (see Tables~\ref{tbl:optimizationImpact} and~\ref{tbl:optimizationImpactDirected} in Appendix~\ref{app:optimizations}).

Tables~\ref{tbl:speedupsDeletionsDirected} and~\ref{tbl:speedupsDeletionsUndirected} in Appendix~\ref{app:complex} show the results for edge deletions (on directed and undirected graphs, respectively). Interestingly, deletions are mostly faster than insertions for directed graphs, whereas they are usually slower in the undirected case. 
For most shortest-paths based problems insertions are easier than deletions: for example, pairwise distances can be updated in time $O(n^2)$ after an edge insertion, but not after an edge deletion~\cite{DBLP:journals/jacm/DemetrescuI04}. In our case, we know deletions can only \textit{decrease} centrality. 
Thus, all previous upper bounds on the centralities are still valid and, if none of the top-$k$ nodes is affected, nothing needs to be updated. In insertions, on the contrary, any affected node could increase its centrality and become one of the top-$k$. 
If the number of affected nodes is small (as it is usually the case for directed graphs, see Table~\ref{tbl:optimizationImpactDirected}), it is quite unlikely that a top-$k$ node is among the affected ones. This happens much more often in undirected graphs, where a larger number of nodes is often affected.
As for insertions, the speedups increase with $k$: for directed graphs, the geometric mean of the speedups is $74$ for $k=1$, $160$ for $k=10$ and $314$ for $k=100$, whereas for undirected graphs it is $12.5$ for $k=1$, $25.8$ for $k=10$ and $50.2$ for $k=100$. All detailed running times for both insertions and deletions can be found in Appendix~\ref{app:time-complex}.

\paragraph{Dynamic Street Networks.} 
The four tables in Appendix~\ref{app:street} show the speedups for street networks, for both edge insertions and deletions, on directed and undirected graphs.
Figure~\ref{fig:summary} summarizes all results for both complex and street networks (the results for street networks are in the upper part of the figure). As for complex networks, speedups in street networks are considerably higher in the directed case. However, differently from complex networks, speedups generally decrease as $k$ increases. In this respect, notice that the running times of the static algorithm (Appendix~\ref{app:time-street}) do not change considerably for different values of $k$ (at least, compared to most complex networks). On the other hand, if $k$ is larger, it is also more likely that some of the affected nodes is either among the top-$k$ or overtakes one of the top-$k$, slowing down the dynamic algorithm. Nevertheless, even for $k=100$, the dynamic algorithm is on average $\approx 49$ times faster than recomputation for insertions in undirected street networks and $\approx 242$ times in directed street networks. The results for deletions are even better: $\approx 67$ in the undirected case and $\approx 519$ in the directed one. The results are significantly better for $k=1$, reaching an average speedup of $\approx 848$ for edge deletions in directed graphs, and $\approx 187$ in undirected graphs.

\begin{figure}[tb]
\begin{center}
\includegraphics[width=17cm]{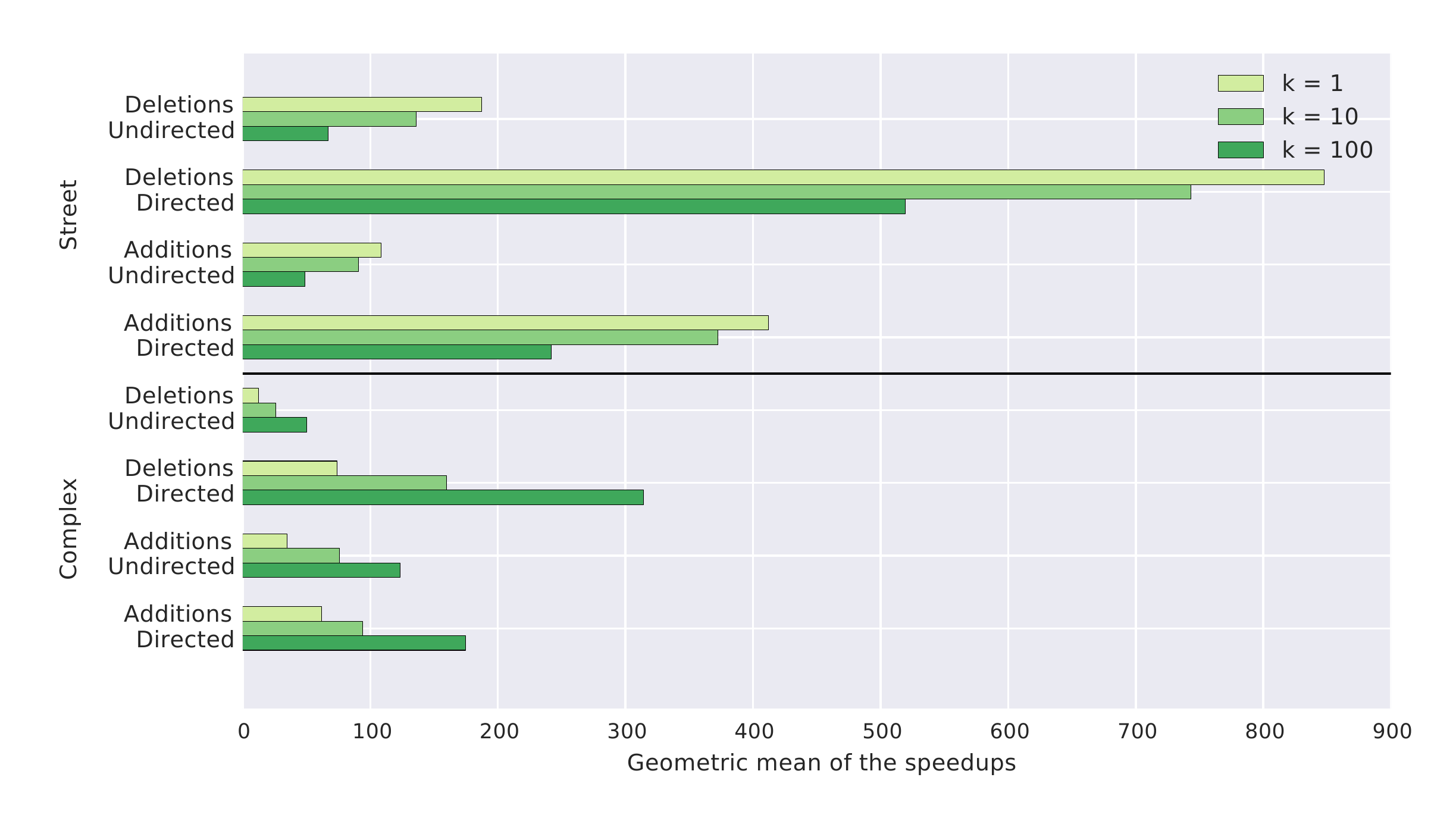}
\caption{Geometric mean of the average speedups over all tested networks, for different values of $k$. The upper part of the plot shows the results for street networks, whereas the lower part shows the results for complex networks. Detailed numbers can be found in Appendix~\ref{app:complex} and Appendix~\ref{app:street}.}
\label{fig:summary}
\end{center}
\vspace{-4ex}
\end{figure}

\section{Conclusions}
\label{sec:conclusions}
We have developed and implemented fully-dynamic algorithms for top-$k$ closeness centrality, tailored to both complex and street networks. Using properties of the modified graph, we are able to significantly reduce the number of operations required to update the most central nodes.
%

As a result, we achieve high speedups on static recomputation, in line with those obtained by other dynamic algorithms for related problems 
\ifthenelse{\boolean{double_blind}}
{
(e.g.,~\cite{DBLP:conf/asunam/KasCC13,DBLP:conf/socialcom/GreenMB12,DBLP:journals/im/BergaminiM16})}
{
(e.g.~\cite{DBLP:conf/asunam/KasCC13,DBLP:journals/im/BergaminiM16,DBLP:conf/wea/BergaminiMOS17,DBLP:conf/socialcom/GreenMB12})}, confirming the fact that efforts in developing dynamic algorithms are well spent. 
Differently from most existing algorithms updating shortest-paths-based centralities, the techniques we propose use a linear (in the number of nodes) amount of additional memory. Although storing more information (e.g., the distances computed during \textsf{BFScut} on the initial graph) might lead to even higher speedups, a quadratic memory footprint would not allow us to target networks with millions of nodes. 
An interesting question is whether the memory requirements of other dynamic algorithms for related problems can be reduced, using similar techniques as the ones presented in this paper.

Future work includes extension to \textit{batch updates}, where several edge updates occur at the same time, and to other centrality measures, such as betweenness. For the latter, a static algorithm for finding the top-$k$ nodes with highest betweenness has been proposed in~\cite{DBLP:conf/www/LeeC14}. Thus, an interesting research question is whether this can be further improved and/or efficiently updated in dynamic networks.


\newpage

\bibliographystyle{abbrv}
\bibliography{references,thesis}

\newpage
\appendix
\section{Pseudocodes}
\label{app:pseudocodes}

\begin{algorithm}[h]
\LinesNumbered
\SetKwFunction{BFS}{BFSbound}
\SetKwFunction{computeBounds}{computeBounds}
\SetKwFunction{extractMax}{extractMax}
\SetKwData{farn}{$c$}
\SetKwData{Q}{Q}
\SetKwFunction{computeReachable}{computeReachable}
\SetKwData{isExact}{isExact}
\SetKwFunction{BFScut}{BFScut}
\SetKwData{L}{$\tilde{c}$}
\SetKwData{Top}{TopK}
 \SetKwInOut{Input}{Input}
 \SetKwInOut{Output}{Output}
\Input{A graph $G=(V,E)$}
\Output{Top-$k$ nodes with highest closeness centrality}
$\Top \leftarrow [\ ]$\;
$r \gets \computeReachable{G}$\; \label{line:reach}
Define an ordering $O$ on the nodes\; \label{line:sort}
$x_k \gets 0$\;
\ForEach{$v \in V$ \text{according to the ordering $O$}}{
	$(\tilde{c}(v), \isExact{v}, d_{cut}(v)) \gets $ \BFScut{$v, x_k$}\; \label{line:bfscut}
	  \If{$\isExact(v) \land \tilde{c}(v) > x_k$}{
  	   \texttt{\Top.insert($\tilde{c}(v)$, $v$)}\;
  	   \If{$\texttt{\Top.size()} > k$}{
	      \texttt{\Top.removeMin()}\;
	   }
	   \If{$\texttt{\Top.size()} = k$}{
	     $x_k \gets \texttt{\Top.getMin()}$\;
	   }
	}
}
\Return{$\Top$}\;

\caption{\textsf{NBCut} algorithm for top-$k$ closeness centrality in static graphs~\cite{DBLP:journals/corr/BergaminiBCMM17}.}
\label{algo:main2}
\end{algorithm}

\begin{algorithm}[h]
\LinesNumbered
\SetKwFunction{BFS}{BFSbound}
\SetKwFunction{computeBounds}{computeBounds}
\SetKwFunction{extractMax}{extractMax}
\SetKwData{farn}{$c$}
\SetKwData{Q}{Q}
\SetKwData{L}{$\tilde{c}$}
\SetKwData{Top}{TopK}
 \SetKwInOut{Input}{Input}
 \SetKwInOut{Output}{Output}
\Input{A graph $G=(V,E)$}
\Output{Top-$k$ nodes with highest closeness centrality}
$\L,\Q \gets \computeBounds{G}$\; \label{line:lbounds}
$\Top \leftarrow [\ ]$\;
\lFor{$v \in V$}{$\farn(v)= 0$}
\While{$\Q$ is not empty} { \label{line:while-start}
    $v \gets \Q.\extractMax()$\;
    \lIf{$|\Top|\geq k$ and $\L(v) > \Top[k]$}{\Return{$\Top$}} \label{line:return}
    $\farn(v) \gets \BFS(v)$; // This function might also modify $\L$ \\
    add $v$ to \Top, and sort \Top according to \farn\;
    update $\Q$ according to the new bounds\;
} \label{line:while-end}
\caption{\textsf{NBBound} algorithm for top-$k$ closeness centrality in static graphs~\cite{DBLP:journals/corr/BergaminiBCMM17}.}
\label{alg:main}
\end{algorithm}

\begin{algorithm}[h!]
 \KwData{$G = (V, E), (u, v) \notin E$}
 \KwResult{Top-$k$ nodes with the highest closeness in $G' := (V, E \cup \{u, v\})$}
Compute $r'(y)\quad \forall y \in V$\; \label{line:preproc1}
 Compute the set $A$ of affected nodes, $d(\cdot, u)$, $d'(\cdot, u)$\; \label{line:preproc2} 
 \SetKwData{Top}{TopK}
  \SetKwData{newBfs}{newBFS}
    \SetKwData{isExact}{isExact}
        \SetKwFunction{BFScut}{BFScut}
 
 $x_k \gets \texttt{\Top.getMin()}$\; \label{line:removeStart}
 \ForAll{$w \in A$}{
 	\If{$w \in \Top$}{
 	   \texttt{\Top.remove($w$)}\;
 	}
 }\label{line:removeEnd}

 \ForEach{$y \in A$}{ \label{alg:dynamicInsertionLoop}

    \uIf{$d_{cut}(y) < d(y, u)\  \land$ \textbf{not} $\isExact(y)$}{ \label{line:opt1}
        \tcc{$y$ is a far-away node}
        $\tilde{c}'(y) \gets \tilde{c}(y) - \frac{r(y)}{d_{cut}(y)+2} + \frac{r'(y)}{d_{cut}(y)+2}$\; \label{line:faraway}
    }
    \uElseIf{$d_{cut}(y) == d(y, u)\  \land$ \textbf{not} $\isExact(y)$}{
        \tcc{$y$ is a boundary node}
      $\tilde{c}'(y) \gets \tilde{c}(y) - \frac{r(y) - r'(y) + 1}{d_{cut}(y) + 2} + \frac{1}{d_{cut}(y) + 1}$\; \label{line:boundary}
      
      }
         \uElse{
        \tcc{we compute the distance-based bounds} 
      $\tilde{c}'(y) \gets \tilde{c}(y) +  \sum_{i = 1}^{\mathsf{diam}} \frac{1}{i + d(y, u)}  \left( {n}'_i(u) -n_i(u) \right)$\; \label{line:dbbound1}
      $d_{cut}(y) \gets \mathsf{diam}$\; \label{line:dbbound2}
    } \label{line:opt2}
    
       \If{$\tilde{c}'(y) \geq x_k$}{    \label{line:dynbfscut1}
    	\tcc{we have to run a new \textsf{BFScut}}
      	$(\tilde{c}'(y), \isExact{y}, d_{cut}(y)) \gets $ \BFScut{$y, x_k$}\; 

  	\If{$\isExact(y) \land \tilde{c}'(y) > x_k$}{
  	   \texttt{\Top.insert($\tilde{c}'(y)$, $y$)}\;
  	   \If{$\texttt{\Top.size()} > k$}{
	      \texttt{\Top.removeMin()}\;
	   }
	   \If{$\texttt{\Top.size()} = k$}{
	     $x_k \gets \texttt{\Top.getMin()}$\;
	   }
	}
} \label{line:dynbfscut2}
 }
 Set $\tilde{c}$, $r$, $n_i$ to $\tilde{c}'$, $r'$, $n'_i$\;
 \caption{Recomputation of the top-$k$ nodes after an edge insertion (based on \textsf{NBCut}).}
  \label{algo:insertions}
\end{algorithm}

\begin{algorithm}[h!]
 \KwData{$G = (V, E), (u, v) \notin E$}
 \KwResult{Top-$k$ nodes with the highest closeness in $G' := (V, E \cup \{u, v\})$}
 \SetKwData{Top}{TopK}
 \SetKwData{Q}{Q}
  \SetKwData{newBfs}{newBFS}
    \SetKwData{isExact}{isExact}
        \SetKwFunction{BFScut}{BFScut}
  Compute the set $A$ of affected nodes, $d(\cdot, u)$, $d'(\cdot, u)$\; 
 \ForAll{$w \in A$}{
 	\If{$w \in \Top$}{
 	   \texttt{\Top.remove($w$)}\;
 	}
 }
 $\Q \gets \emptyset $\;
 \ForEach{$y \in A$}{ 
        \tcc{we update the bounds on the affected nodes} 
      $\tilde{c}'(y) \gets \tilde{c}(y) +  \sum_{i = 1}^{\mathsf{diam}} \frac{1}{i + d(y, u)}  \left( {n}'_i(u) -n_i(u) \right)$\; 
      insert $y$ into $\Q$ with priority $\tilde{c}'(y)$\;
}

\While{$\Q$ is not empty} { \label{line:while-start2}
    $y \gets \Q.\extractMax()$\;
    \lIf{$|\Top|\geq k$ and $\tilde{c}(y) > \Top[k]$}{\Return{$\Top$}} \label{line:return2}
    $c(y) \gets \BFS(v)$; // This function might also modify $\tilde{c}$ \\
    add $y$ to \Top, and sort \Top according to $c$\;
    update $\Q$ according to the new bounds\;
} \label{line:while-end2}
 \caption{Recomputation of the top-$k$ nodes after an edge insertion (based on \textsf{NBBound}).}
  \label{algo:insertions2}
\end{algorithm}

\begin{algorithm}[h!]
 \KwData{$G = (V, E), (u, v) \notin E$}
 \KwResult{Top-$k$ nodes with the highest closeness in $G' := (V, E \cup \{u, v\})$}
 \SetKwInOut{Assume}{Assume}
 \Assume{$Q = $ priority queue with nodes sorted by $\tilde{c}$\;}
Compute $r'(y)\quad \forall y \in V$\; \label{line:preproc1d}
 Compute the set $A$ of affected nodes, $d(\cdot, u)$, $d'(\cdot, u)$\;  
 \SetKwData{Top}{TopK}
\SetKwData{newBfs}{newBFS}
\SetKwData{isExact}{isExact}
\SetKwFunction{BFScut}{BFScut}
\SetKwData{farn}{$c$}
\SetKwData{Q}{Q}
\SetKwData{L}{$\tilde{c}$}
 
 \ForAll{$w \in A$}{
 	\If{$w \in \texttt{\Top}$}{
 	   \texttt{\Top.remove($w$)}\;
 	}
	$\isExact(w) \gets $ false\;
 }
 $x_k \gets 0$\; 
 
 \While{$\Q$ is not empty} { 
    $y \gets \Q.\extractMax()$\;
    \lIf{$|\Top|\geq k$ and $\L(y) > \Top[k]$}{\Return{$\Top$}} \label{line:return}
    \If{not $\isExact(y)$}{
    \tcc{We run a new \textsf{BFScut}}
    	$(\tilde{c}'(y), \isExact{y}, d_{cut}(y)) \gets $ \BFScut{$y, x_k$}\; \label{line:bfscut-del}
    }
      	\If{$\isExact(y) \land \tilde{c}'(y) > x_k$}{
  	   \texttt{\Top.insert($\tilde{c}'(y)$, $y$)}\;
  	   \If{$\texttt{\Top.size()} > k$}{
	      \texttt{\Top.removeMin()}\;
	   }
	   \If{$\texttt{\Top.size()} = k$}{
	     $x_k \gets \texttt{\Top.getMin()}$\;
	   }
	}

}  
 Set $\tilde{c}$, $r$, $n_i$ to $\tilde{c}'$, $r'$, $n'_i$\;
 \caption{Recomputation of the top-$k$ nodes after an edge deletion (based on \textsf{NBCut}).}
  \label{algo:deletions}
\end{algorithm}

\FloatBarrier
\section{Additional experimental results}
\label{app:tables}
\setlength{\tabcolsep}{6pt}
\subsection{Overview of Networks}
\begin{table}[H]
\centering
\begin{footnotesize}
\begin{minipage}{0.48\textwidth}
\centering
\begin{tabular}{lrr}
\toprule
 Graph            &   Nodes &   Edges \\
\midrule \midrule
 polblogs         &    1224 &   19025 \\
 p2p-Gnutella08   &    6301 &   20777 \\
 wiki-Vote        &    7115 &  103689 \\
 p2p-Gnutella09   &    8114 &   26013 \\
 p2p-Gnutella06   &    8717 &   31525 \\
 p2p-Gnutella04   &   10876 &   39994 \\
 freeassoc        &   14213 &   72176 \\
 as-caida20071105 &   26475 &  106762 \\
 p2p-Gnutella31             &   62586 &    147892 \\
 soc-Epinions1              &   75879 &    508837 \\
 web-Stanford               &  281903 &   2312497 \\
 web-NotreDame              &  325729 &   1469679 \\
 wiki-Talk      			& 2394385  & 5021410 \\
 cit-Patents                & 3774768 &  16518948 \\
\bottomrule
\end{tabular}
\caption{Overview of directed complex networks  used in the experiments.}
\label{tbl:datasetDirected}
\end{minipage}
\begin{minipage}{0.48\textwidth}
\centering
\begin{tabular}{lrr}
\toprule
 Graph                   &   Nodes &   Edges \\
\midrule \midrule
HC-BIOGRID              &    4039 &   10321 \\
Mus\_musculus            &    4610 &    5747 \\
Caenorhabditis\_elegans  &    4723 &    9842 \\
ca-GrQc                 &    5241 &   14484 \\
advogato                &    7418 &   48037 \\
hprd\_pp                 &    9465 &   37039 \\
ca-HepTh                &    9877 &   25998 \\
Drosophila\_melanogaster &   10625 &   40781 \\ 
oregon1\_010526          &   11174 &   23409 \\
oregon2\_010526 & 11461 &   32730 \\
Homo\_sapiens            &   13690 &   61130 \\
GoogleNw                &   15763 &  148585 \\
dip20090126\_MAX         &   19928 &   41202 \\
ca-HepPh                &   12008 &    118521 \\
CA-AstroPh              &   18772 &    198110 \\
CA-CondMat              &   23133 &     93497 \\
email-Enron             &   36692 &    183831 \\
loc-brightkite    &   58228 &    214078 \\
gowalla           &  196591 &    950327 \\
com-dblp        &  317080 &   1049866 \\
com-amazon      &  334863 &    925872 \\
com-youtube & 1134890 & 2987624 \\
\bottomrule
\end{tabular}
\caption{Overview of undirected complex networks  used in the experiments.}
\label{tbl:datasetUndirected}
\end{minipage}
\end{footnotesize}
\end{table}
\begin{table}[H]
\centering
\begin{footnotesize}
\begin{tabular}{lrrr}
\toprule
 Graph         &   Nodes &   Edges \\
\midrule \midrule
 maldives                     &    22591 &    27088 \\
 faroe-islands &   31097 &   31974 \\
 liechtenstein &   54972 &   56616  \\
 isle-of-man   &   61082 &   63793  \\
 equatorial-guinea            &    80262 &    82245 \\
 malta         &   91188 &  101437  \\
 belize        &   96977 &  103198 \\
 azores        &  237174 &  243185 \\
\bottomrule
\end{tabular}
\end{footnotesize}
\caption{Overview of street networks used in the experiments.}
\label{tbl:streetNetworks}
\end{table}

\FloatBarrier
\subsection{Impact of Optimizations}
\label{app:optimizations}
\begin{table}[H]
\centering
\begin{footnotesize}
\begin{tabular}{lrrrrrr}
\toprule
 Graph            &   affected &   \% affected & far-away   & boundary   & dist. bound   & \textsf{BFScuts}   \\
\midrule \midrule
polblogs         &         39 & 3.166\%  & 41.084\%    & 13.265\%    & 36.310\%       & 9.342\%   \\
 p2p-Gnutella08   &        311 & 4.942\%  & 42.941\%    & 19.577\%    & 6.102\%        & 31.380\%  \\
 wiki-Vote        &         31 & 0.440\%  & 0.287\%     & 0.255\%     & 79.355\%       & 20.102\%  \\
 p2p-Gnutella09   &        426 & 5.245\%  & 50.895\%    & 17.709\%    & 12.175\%       & 19.220\%  \\
 p2p-Gnutella06   &        589 & 6.752\%  & 60.853\%    & 20.203\%    & 7.343\%        & 11.601\%  \\
 p2p-Gnutella04   &        773 & 7.108\%  & 67.576\%    & 17.189\%    & 6.639\%        & 8.597\%   \\
 freeassoc        &        323 & 2.273\%  & 0.000\%     & 0.000\%     & 95.220\%       & 4.780\%   \\    
 as-caida20071105 &       2076 & 7.843\%  & 87.803\%    & 11.912\%    & 0.247\%        & 0.038\%   \\
 p2p-Gnutella31   &       2895 & 4.625\%  & 71.353\%    & 13.087\%    & 10.360\%       & 5.200\%   \\
 soc-Epinions1    &        659 & 0.868\%  & 13.611\%    & 38.786\%    & 20.705\%       & 26.898\%  \\
 wiki-Talk        &       3411 & 0.142\%  & 96.579\%    & 3.306\%     & 0.106\%        & 0.009\%   \\
 web-Stanford     &      10968 & 3.891\%  & 66.812\%    & 12.232\%    & 14.690\%       & 6.265\%   \\
 web-NotreDame    &       3896 & 1.196\%  & 94.599\%    & 2.200\%     & 3.112\%        & 0.090\%   \\
  cit-Patents      &        161 & 0.004\%  & 0.000\%     & 0.000\%     & 99.901\%       & 0.099\%   \\
\bottomrule
\end{tabular}
\end{footnotesize}
\caption{Impact of optimizations in directed networks for $k = 10$, averaged over 100 random edge insertions. The column ``affected'' contains the average number of affected nodes and ``\% affected'' its percentage of the total number of nodes. The next three columns report the percentage of affected nodes that can be skipped for each optimization and the last column the percentage of affected nodes for which a \textsf{BFScut} has been run.}
\label{tbl:optimizationImpactDirected}
\end{table}

\begin{table}[H]
\centering
\begin{footnotesize}
\begin{tabular}{lrrrrrr}
\toprule
 Graph                &   affected &   \% affected & far-away   & boundary   & dist. bound   & \textsf{BFScuts}   \\
\midrule \midrule
 HC-BIOGRID              &        598 & 14.812\% & 61.555\%    & 8.876\%     & 27.527\%       & 2.043\%   \\
 Mus\_musculus            &       1556 & 33.762\% & 62.127\%    & 7.547\%     & 29.682\%       & 0.644\%   \\
 Caenorhabditis\_elegans  &        508 & 10.758\% & 70.768\%    & 6.794\%     & 21.452\%       & 0.986\%   \\
 ca-GrQc                 &        690 & 13.170\% & 81.017\%    & 7.090\%     & 10.792\%       & 1.101\%   \\  
 advogato                &        281 & 3.784\%  & 58.696\%    & 18.037\%    & 21.496\%       & 1.771\%   \\
 hprd\_pp                 &        731 & 7.728\%  & 92.468\%    & 5.731\%     & 1.471\%        & 0.329\%   \\
 ca-HepTh                &       1362 & 13.795\% & 89.968\%    & 5.780\%     & 2.641\%        & 1.611\%   \\
 Drosophila\_melanogaster &        836 & 7.872\%  & 85.411\%    & 10.178\%    & 2.483\%        & 1.927\%   \\
 oregon1\_010526          &       1932 & 17.288\% & 76.625\%    & 22.486\%    & 0.805\%        & 0.084\%   \\
 oregon2\_010526          &       1470 & 12.824\% & 76.226\%    & 22.781\%    & 0.774\%        & 0.220\%   \\
 Homo\_sapiens            &        692 & 5.054\%  & 93.008\%    & 5.686\%     & 1.003\%        & 0.304\%   \\
 GoogleNw                &        838 & 5.316\%  & 39.976\%    & 59.204\%    & 0.717\%        & 0.103\%   \\
 dip20090126\_MAX         &       3082 & 15.467\% & 44.315\%    & 11.963\%    & 35.153\%       & 8.569\%   \\      
 ca-HepPh              &        745 & 6.204\%  & 77.089\%    & 3.473\%     & 18.230\%       & 1.208\%   \\
 CA-AstroPh              &        795 & 4.233\%  & 93.756\%    & 4.596\%     & 1.506\%        & 0.142\%   \\
 CA-CondMat              &       3026 & 13.081\% & 93.658\%    & 5.090\%     & 1.050\%        & 0.202\%   \\
 email-Enron             &       2097 & 5.716\%  & 93.122\%    & 6.144\%     & 0.665\%        & 0.069\%   \\
 loc-brightkite    &       5973 & 10.259\% & 72.360\%    & 7.362\%     & 19.820\%       & 0.459\%   \\
 gowalla           &      16581 & 8.434\%  & 74.731\%    & 1.456\%     & 23.786\%       & 0.026\%   \\
 com-dblp        &      67781 & 21.377\% & 92.218\%    & 2.708\%     & 5.044\%        & 0.029\%   \\
 com-amazon      &      71059 & 21.220\% & 88.512\%    & 3.888\%     & 7.057\%        & 0.543\%   \\
 com-youtube     &     101963 & 8.984\%  & 86.257\%    & 12.424\%    & 0.908\%        & 0.411\%   \\
\bottomrule
\end{tabular}
\end{footnotesize}
\caption{Impact of optimizations in undirected complex networks for $k = 10$, averaged over 100 random edge insertions. The column ``affected'' contains the average number of affected nodes and ``\% affected'' its percentage of the total number of nodes. The next three columns report the percentage of affected nodes that can be skipped for each optimization and the last column the percentage of affected nodes for which a \textsf{BFScut} has been run.}
\label{tbl:optimizationImpact}
\end{table}

\FloatBarrier

\subsection{Speedups for Complex Networks}
\label{app:complex}
\begin{table}[H]
\centering
\begin{footnotesize}
\begin{tabular}{lrrr|rrr|rrr}
\toprule
Graph	&  \multicolumn{3}{c}{k = 1} & \multicolumn{3}{c}{k = 10} & \multicolumn{3}{c}{k = 100}\\
            &    gmean &     min &      max &    gmean &    min &      max &    gmean &     min &      max \\
\midrule \midrule
  polblogs         &    41.6 &   11.3 &    67.5 &    78.5 &   1.9 &   149.2 &   146.8 &   3.5 &   252.4 \\
 p2p-Gnutella08   &    18.9 &    1.7 &    34.1 &    32.6 &   0.7 &   268.6 &    53.7 &   1.2 &   503.1 \\
 wiki-Vote        &    52.6 &    2.0 &    85.1 &    53.1 &   1.2 &   156.0 &   191.4 &   7.2 &   298.2 \\
 p2p-Gnutella09   &    37.8 &    5.8 &    75.5 &    24.7 &   1.0 &   189.3 &    45.7 &   1.3 &   707.7 \\
 p2p-Gnutella06   &    21.4 &    9.4 &    34.7 &    26.0 &   1.4 &   210.2 &    45.8 &   1.8 &   689.9 \\
 p2p-Gnutella04   &    23.9 &    9.0 &    35.3 &    30.6 &   2.2 &   277.6 &    34.4 &   2.4 &   804.3 \\
 freeassoc        &   433.8 &   91.2 &   685.8 &   347.4 &   1.6 &   766.1 &   403.1 &  11.8 &   831.6 \\
 as-caida20071105 &     9.6 &    2.7 &    12.7 &    14.1 &   2.1 &    19.5 &    99.3 &   3.9 &   221.2 \\
 p2p-Gnutella31   &    76.1 &   16.4 &   194.2 &    48.0 &   3.1 &   491.8 &    21.2 &   1.1 &  1084.7 \\
 soc-Epinions1    &    11.1 &    0.1 &    73.8 &    35.1 &   0.1 &   386.7 &    84.7 &   0.3 &  1284.7 \\
 web-Stanford     &  3367.6 &    0.4 & 21500.8 &  3364.7 &   0.6 & 28577.8 &  5270.0 &   0.9 & 41562.7 \\
 web-NotreDame    &    50.7 &   12.5 &    72.3 &   569.7 &   6.0 &   901.6 &  1363.2 &  15.2 &  2560.3 \\
 wiki-Talk        &    14.5 &   11.1 &    15.7 &    43.6 &   6.0 &    51.2 &   174.3 &  11.1 &   235.0 \\
 cit-Patents      &  1699.1 & 1127.4 &  1897.9 &  2237.3 & 823.2 &  2636.3 &  2513.0 & 365.6 &  3393.6 \\
 \midrule \midrule
 (geometric) mean &    61.9 &    6.8 &   118.2 &    93.9 &   2.3 &   362.7 &   174.7 &   3.8 &   922.6 \\ \bottomrule
\end{tabular}
\end{footnotesize}
\caption{Speedups on recomputation over 100 edge insertions in directed complex networks, for $k \in \{1, 10, 100\}$. The column ``gmean'' contains the geometric mean of the achieved speedups, ``min'' and ``max'' the minimum and the maximum speedup.}
\label{tbl:speedupsInsertionDirected}
\end{table}

\begin{table}[H]
\centering
\begin{footnotesize}
\begin{tabular}{lrrr|rrr|rrr}
\toprule
Graph	&  \multicolumn{3}{c}{k = 1} & \multicolumn{3}{c}{k = 10} & \multicolumn{3}{c}{k = 100}\\
           &   gmean &    min &      max &   gmean &    min &      max &   gmean &    min &      max \\
\midrule \midrule
  HC-BIOGRID              &    37.8 &   6.1 &   65.3 &    42.0 &   6.2 &   97.0 &    79.0 &   4.9 &  332.4 \\
 Mus\_musculus            &     8.4 &   3.6 &   39.6 &    14.0 &   3.7 &  101.8 &    20.5 &   2.6 &  358.4 \\
 Caenorhabditis\_elegans  &    12.1 &   3.7 &   17.5 &    18.9 &   3.7 &   35.9 &    35.7 &   2.9 &  162.6 \\
 ca-GrQc                 &    25.9 &   8.5 &   84.3 &    39.7 &   7.2 &  159.7 &    53.4 &   3.4 &  420.9 \\
 advogato                &    37.5 &  11.8 &   42.6 &    62.1 &   2.6 &  108.2 &   107.4 &   3.2 &  268.3 \\
 hprd\_pp                 &    18.1 &   4.4 &   26.5 &    31.8 &   3.6 &   55.2 &    51.8 &   3.6 &  323.1 \\
 ca-HepTh                &    30.8 &   3.6 &  205.9 &    36.8 &   4.7 &  385.0 &    54.7 &   4.4 &  957.5 \\
 Drosophila\_melanogaster &    19.3 &   5.3 &   27.0 &    59.6 &   9.5 &  156.1 &    60.2 &   3.5 &  272.6 \\
 oregon1\_010526          &    12.2 &   2.9 &   19.4 &    15.1 &   2.4 &   29.5 &    50.8 &   2.6 &  225.4 \\
 oregon2\_010526          &    16.7 &   3.8 &   24.1 &    22.4 &   2.9 &   44.8 &    80.8 &   3.4 &  258.8 \\
 ca-HepPh                &   365.4 &  15.6 &  704.1 &   397.6 &   3.7 & 3352.4 &   487.0 &   7.1 &  972.2 \\
 Homo\_sapiens            &    20.6 &   5.5 &   26.6 &    34.8 &   6.2 &   64.2 &    59.1 &   4.8 &  424.8 \\
 GoogleNw                &    31.3 &   8.8 &   36.9 &   798.4 &  78.8 & 1045.3 &   886.6 &  11.4 & 1452.1 \\
 CA-AstroPh              &   133.1 &   6.8 &  288.1 &   280.1 &  28.5 &  414.5 &   418.4 &  11.3 & 6602.3 \\
 dip20090126\_MAX         &    53.1 &   6.5 &  645.1 &    54.1 &   6.2 & 1096.5 &    57.1 &   9.3 & 1878.0 \\
 CA-CondMat              &    35.8 &   9.3 &  183.8 &    50.1 &   5.0 &  417.8 &    94.7 &   3.3 & 1458.5 \\
 email-Enron             &    56.0 &   2.0 &  267.2 &   105.9 &  11.0 &  442.8 &   222.6 &   5.2 &  446.8 \\
 loc-brightkite    &    23.2 &   4.3 &   33.1 &   133.2 &   2.9 & 1200.6 &   132.9 &   5.0 & 2028.6 \\
 gowalla           &    15.9 &   3.8 &   22.1 &   142.1 &   8.4 &  257.3 &   936.0 &   6.7 & 3463.2 \\
 com-dblp        &   140.1 &  17.6 &  282.7 &   130.4 &  15.1 &  357.3 &   108.7 &   4.7 &  766.4 \\
 com-amazon      &   445.1 &  19.6 & 1389.8 &   304.6 &   8.8 & 1853.6 &   518.3 &   9.0 & 5538.1 \\
 com-youtube     &    11.6 &   2.8 &   17.0 &   505.9 &   4.1 & 2122.3 &   479.2 &   2.0 & 3669.0 \\
  \midrule \midrule
 (geometric) mean        &    34.6 &   5.9 &   77.3 &    75.7 &   6.3 &  265.9 &   123.3 &   4.6 &  790.7 \\ \bottomrule
\end{tabular}
\end{footnotesize}
\caption{Speedups on recomputation over 100 edge insertions in undirected complex networks, for $k \in \{1, 10, 100\}$. The column ``gmean'' contains the geometric mean of the achieved speedups, ``min'' and ``max'' the minimum and the maximum speedup.}
\label{tbl:speedupsInsertionUndirected}
\end{table}

\begin{table}[H]
\centering
\begin{footnotesize}
\begin{tabular}{lrrr|rrr|rrr}
\toprule
Graph	&  \multicolumn{3}{c}{k = 1} & \multicolumn{3}{c}{k = 10} & \multicolumn{3}{c}{k = 100}\\
           &   gmean &    min &      max &   gmean &    min &      max &   gmean &    min &      max \\
\midrule \midrule 
 polblogs         &    43.0 &   29.7 &   47.7 &    94.7 &   78.4 &   102.5 &   106.6 &   61.2 &    151.3 \\
 p2p-Gnutella08   &    16.6 &    2.0 &   31.6 &    59.2 &    0.8 &   229.6 &   114.2 &    2.5 &    523.9 \\
 wiki-Vote        &    32.1 &   28.6 &   33.9 &    61.2 &   59.1 &    63.3 &    98.2 &   32.4 &    116.7 \\
 p2p-Gnutella09   &    36.6 &    8.4 &   75.2 &    36.1 &    1.1 &   182.4 &    35.7 &    4.3 &    155.7 \\
 p2p-Gnutella06   &    45.0 &   13.4 &   53.4 &    33.4 &    1.6 &   148.9 &   145.0 &    4.8 &    614.4 \\
 p2p-Gnutella04   &    23.5 &    8.9 &   27.4 &    63.7 &    5.2 &   236.2 &    89.7 &    7.7 &    794.0 \\
 freeassoc        &   373.5 &  305.6 & 1412.8 &   397.8 &  331.3 &  1398.6 &   332.1 &   72.2 &   1683.0 \\
 as-caida20071105 &     8.6 &    8.1 &    8.8 &    14.0 &   12.1 &    14.5 &   154.2 &  124.5 &    161.9 \\
 p2p-Gnutella31   &    68.8 &   10.3 &  220.0 &   132.5 &    2.9 &   652.2 &   190.4 &    7.5 &   1585.5 \\
 soc-Epinions1    &    41.1 &   37.0 &   43.4 &   138.6 &  108.7 &   165.0 &   629.3 &  266.8 &    754.6 \\
 web-Stanford     &  3796.1 & 3711.1 & 3887.1 &  6019.9 & 4281.5 & 55778.7 &  9471.6 & 6989.9 & 104810.2 \\
 web-NotreDame    &    58.7 &   49.6 &   69.6 &   859.7 &  781.9 &   896.5 &  2353.9 & 2166.3 &   2486.6 \\
 wiki-Talk        &    21.6 &   20.1 &   22.8 &    81.1 &   75.8 &    84.8 &   348.4 &  335.9 &    366.2 \\
 cit-Patents      &  3717.3 & 2923.6 & 4394.9 &  6357.6 & 3543.7 &  7096.1 &  3472.4 &  180.5 &   7523.4 \\
 \midrule \midrule
 (geometric) mean &    73.9 &   38.8 &  104.3 &   159.9 &   40.9 &   360.5 &   314.3 &   62.2 &    879.0 \\ \bottomrule
 \end{tabular}
\end{footnotesize}
\caption{Speedups on recomputation over 100 edge removals in directed complex networks, for $k \in \{1, 10, 100\}$. The column ``gmean'' contains the geometric mean of the achieved speedups, ``min'' and ``max'' the minimum and the maximum speedup.}
\label{tbl:speedupsDeletionsDirected}
\end{table}

\begin{table}[H]
\centering
\begin{footnotesize}
\begin{tabular}{lrrr|rrr|rrr}
\toprule
Graph	&  \multicolumn{3}{c}{k = 1} & \multicolumn{3}{c}{k = 10} & \multicolumn{3}{c}{k = 100}\\
           &   gmean &    min &      max &   gmean &    min &      max &   gmean &    min &      max \\
\midrule \midrule 
 HC-BIOGRID              &     8.8 &   1.8 &   15.2 &    11.3 &   1.3 &   24.6 &    24.2 &   2.4 &    86.8 \\
 Mus\_musculus            &     6.0 &   2.9 &  104.8 &     9.6 &   2.8 &  350.2 &    13.9 &   2.2 &   679.9 \\
 Caenorhabditis\_elegans  &     4.2 &   1.6 &    6.3 &     6.1 &   1.5 &   13.2 &    12.9 &   1.8 &    58.8 \\
 ca-GrQc                 &    17.1 &   1.7 &  339.4 &    22.6 &   2.2 &  576.8 &    39.8 &   2.4 &  1679.5 \\
 advogato                &    11.1 &   2.4 &  407.8 &    20.7 &   1.1 & 1207.4 &    35.9 &   1.4 &  3046.0 \\
 hprd\_pp                 &     5.6 &   1.6 &    6.9 &     8.2 &   1.4 &   14.4 &    22.9 &   1.3 &    86.3 \\
 ca-HepTh                &    12.7 &   1.3 &  478.3 &    17.2 &   1.2 &  937.3 &    27.3 &   2.0 &  2436.8 \\
 Drosophila\_melanogaster &     5.6 &   1.6 &    6.8 &    21.3 &   1.4 &   39.2 &    20.3 &   1.7 &    69.7 \\
 oregon1\_010526          &     3.9 &   2.0 &    5.1 &     4.5 &   1.7 &    7.9 &    14.9 &   1.6 &    63.1 \\
 oregon2\_010526          &     4.6 &   2.0 &    5.9 &     6.1 &   1.5 &   11.3 &    20.2 &   2.4 &    66.0 \\
 ca-HepPh                &   116.8 &   1.0 & 8902.1 &   109.4 &   1.0 & 9336.2 &   129.7 &   1.0 & 12759.1 \\
 Homo\_sapiens            &     5.5 &   1.9 &    6.5 &     9.1 &   1.2 &   15.8 &    31.5 &   1.5 &   106.9 \\
 GoogleNw                &     9.1 &   2.3 &   13.0 &   149.5 &   1.0 &  354.1 &   164.7 &   1.1 &   507.9 \\
 CA-AstroPh              &    65.6 &   1.1 & 4020.2 &    88.4 &   1.1 & 5886.6 &   207.8 &   1.1 & 18076.6 \\
 dip20090126\_MAX         &    89.4 &   1.7 &  182.4 &    81.2 &   1.4 &  264.5 &   102.9 &   1.8 &   487.4 \\
 CA-CondMat              &    13.5 &   2.2 &  470.7 &    21.4 &   1.3 &  996.5 &    47.2 &   1.8 &  3807.3 \\
 email-Enron             &    17.1 &   1.3 &  780.5 &    30.1 &   7.6 & 1204.6 &    63.5 &   4.8 &  3607.3 \\
 loc-brightkite    &     6.0 &   1.6 &  230.0 &    40.2 &   1.1 & 2566.3 &    45.6 &   1.4 &  4319.3 \\
 gowalla           &     4.6 &   2.2 &    5.4 &    36.2 &   1.2 &   60.0 &   336.8 &   1.3 &   836.0 \\
 com-dblp        &    37.3 &   2.7 &   81.6 &    39.3 &   2.3 &  101.9 &    50.3 &   1.9 &   219.3 \\
 com-amazon      &   155.9 &   8.1 &  461.2 &    99.4 &   3.0 &  539.6 &   211.3 &   9.5 &  1595.0 \\
 com-youtube     &     3.4 &   1.9 &    4.4 &   117.4 &   1.0 &  585.6 &   156.9 &   1.1 &  1029.9 \\
  \midrule \midrule
 (geometric) mean        &    12.5 &   1.9 &   70.0 &    25.8 &   1.5 &  213.6 &    50.2 &   1.8 &   662.9 \\
 \bottomrule
 \end{tabular}
 \end{footnotesize}
\caption{Speedups on recomputation over 100 edge removals in undirected complex networks, for $k \in \{1, 10, 100\}$. The column ``gmean'' contains the geometric mean of the achieved speedups, ``min'' and ``max'' the minimum and the maximum speedup. }
\label{tbl:speedupsDeletionsUndirected}
\end{table}

\FloatBarrier

\subsection{Running Times for Complex Networks}
\label{app:time-complex}

\begin{table}[H]
\centering
\begin{footnotesize}
\begin{tabular}{lrr|rr|rr}
\toprule
Graph & \multicolumn{2}{c}{k = 1} & \multicolumn{2}{c}{k = 10} & \multicolumn{2}{c}{k = 100}\\
                    &     static [s] &   dynamic [s] &     static[s] &   dynamic [s] &     static [s] &   dynamic [s] \\\midrule \midrule
  polblogs         &   0.0121 &    0.0003 &   0.0298 &    0.0004 &   0.0586 &    0.0004 \\
 p2p-Gnutella08   &   0.0148 &    0.0008 &   0.1026 &    0.0031 &   0.2636 &    0.0049 \\
 wiki-Vote        &   0.0536 &    0.0010 &   0.0969 &    0.0018 &   0.1857 &    0.0010 \\
 p2p-Gnutella09   &   0.0446 &    0.0012 &   0.1072 &    0.0043 &   0.3640 &    0.0080 \\
 p2p-Gnutella06   &   0.0242 &    0.0011 &   0.1293 &    0.0050 &   0.4541 &    0.0099 \\
 p2p-Gnutella04   &   0.0323 &    0.0013 &   0.2238 &    0.0073 &   0.8322 &    0.0242 \\
 freeassoc        &   0.2658 &    0.0006 &   0.2977 &    0.0009 &   0.3493 &    0.0009 \\
 as-caida20071105 &   0.0416 &    0.0043 &   0.0702 &    0.0050 &   0.7898 &    0.0079 \\
 p2p-Gnutella31   &   0.8200 &    0.0108 &   2.4506 &    0.0511 &   6.0270 &    0.2841 \\
 soc-Epinions1    &   0.8373 &    0.0752 &   2.9520 &    0.0840 &  12.7048 &    0.1500 \\
 web-Stanford     & 207.7361 &    0.0617 & 288.4620 &    0.0857 & 406.5858 &    0.0772 \\
 web-NotreDame    &   1.1105 &    0.0219 &  13.9541 &    0.0245 &  40.1074 &    0.0294 \\
 wiki-Talk        &   2.9550 &    0.2035 &   9.6525 &    0.2212 &  43.4118 &    0.2491 \\
 cit-Patents      & 351.6200 &    0.2069 & 371.7316 &    0.1662 & 428.1452 &    0.1704 \\
\bottomrule
\end{tabular}
\end{footnotesize}
\caption{Update times for 100 random edge insertions with  $k \in \{1, 10, 100\}$ in directed complex networks. The columns ``static''  and ``dynamic'' contain the average time for the static and dynamic algorithm, respectively.}
\label{tbl:insertionComplexNetworkDirApp}
\end{table}

\begin{table}[H]
\centering
\begin{footnotesize}
\begin{tabular}{lrr|rr|rr}
\toprule
Graph & \multicolumn{2}{c}{k = 1} & \multicolumn{2}{c}{k = 10} & \multicolumn{2}{c}{k = 100}\\
                   &     static [s] &   dynamic [s] &     static [s] &   dynamic [s] &    static [s] &   dynamic [s] \\\midrule \midrule
  HC-BIOGRID              &   0.0317 &    0.0008 &   0.0527 &    0.0013 &   0.1791 &    0.0023 \\
 Mus\_musculus            &   0.0083 &    0.0010 &   0.0260 &    0.0019 &   0.0976 &    0.0048 \\
 Caenorhabditis\_elegans  &   0.0106 &    0.0009 &   0.0217 &    0.0011 &   0.0980 &    0.0027 \\
 ca-GrQc                 &   0.0270 &    0.0010 &   0.0469 &    0.0012 &   0.1324 &    0.0025 \\
 advogato                &   0.0491 &    0.0013 &   0.1289 &    0.0021 &   0.3229 &    0.0030 \\
 hprd\_pp                 &   0.0339 &    0.0019 &   0.0708 &    0.0022 &   0.4217 &    0.0081 \\
 ca-HepTh                &   0.0705 &    0.0023 &   0.1328 &    0.0036 &   0.3514 &    0.0064 \\
 Drosophila\_melanogaster &   0.0385 &    0.0020 &   0.2217 &    0.0037 &   0.3956 &    0.0066 \\
 oregon1\_010526          &   0.0173 &    0.0014 &   0.0263 &    0.0017 &   0.2057 &    0.0040 \\
 oregon2\_010526          &   0.0233 &    0.0014 &   0.0441 &    0.0020 &   0.2568 &    0.0032 \\
 ca-HepPh                &   1.4801 &    0.0041 &   1.5431 &    0.0039 &   2.0807 &    0.0043 \\
 Homo\_sapiens            &   0.0506 &    0.0025 &   0.1228 &    0.0035 &   0.8286 &    0.0140 \\
 GoogleNw                &   0.0810 &    0.0026 &   2.3284 &    0.0029 &   3.3147 &    0.0037 \\
 CA-AstroPh              &   1.0297 &    0.0077 &   1.4850 &    0.0053 &   4.5794 &    0.0109 \\
 dip20090126\_MAX         &   1.6690 &    0.0314 &   2.4519 &    0.0453 &   4.5685 &    0.0800 \\
 CA-CondMat              &   0.1445 &    0.0040 &   0.3113 &    0.0062 &   1.1810 &    0.0125 \\
 email-Enron             &   0.3866 &    0.0069 &   0.6290 &    0.0059 &   1.9659 &    0.0088 \\
 loc-brightkite    &   0.2249 &    0.0097 &   2.3622 &    0.0177 &   4.3341 &    0.0326 \\
 gowalla           &   0.8203 &    0.0515 &   9.0723 &    0.0638 & 127.8335 &    0.1366 \\
 com-dblp        &  16.8464 &    0.1202 &  20.9283 &    0.1605 &  44.7540 &    0.4117 \\
 com-amazon      & 102.1518 &    0.2295 & 118.0336 &    0.3875 & 345.3928 &    0.6663 \\
 com-youtube     &   3.5802 &    0.3074 & 427.8860 &    0.8458 & 759.7169 &    1.5854 \\  \bottomrule
\end{tabular}
\end{footnotesize}
\caption{Update times for 100 random edge insertions with  $k \in \{1, 10, 100\}$ in undirected complex networks. The columns ``static''  and ``dynamic'' contain the average time for the static and dynamic algorithm, respectively.}
\label{tbl:insertionComplexNetworkUndirApp}
\end{table}

\begin{table}[H]
\centering
\begin{footnotesize}
\begin{tabular}{lrr|rr|rr}
\toprule
Graph & \multicolumn{2}{c}{k = 1} & \multicolumn{2}{c}{k = 10} & \multicolumn{2}{c}{k = 100}\\
           &    static [s] &   dynamic [s] &     static [s] &   dynamic [s] &     static [s] &   dynamic [s] \\
\midrule \midrule
  polblogs         &   0.0188 &    0.0004 &   0.0401 &    0.0004 &   0.0580 &    0.0005 \\
 p2p-Gnutella08   &   0.0163 &    0.0010 &   0.1189 &    0.0020 &   0.2669 &    0.0023 \\
 wiki-Vote        &   0.0527 &    0.0016 &   0.0995 &    0.0016 &   0.1859 &    0.0019 \\
 p2p-Gnutella09   &   0.0448 &    0.0012 &   0.1087 &    0.0030 &   0.3623 &    0.0101 \\
 p2p-Gnutella06   &   0.0444 &    0.0010 &   0.1245 &    0.0037 &   0.5147 &    0.0035 \\
 p2p-Gnutella04   &   0.0307 &    0.0013 &   0.2646 &    0.0042 &   0.8984 &    0.0100 \\
 freeassoc        &   0.2904 &    0.0008 &   0.3084 &    0.0008 &   0.3455 &    0.0010 \\
 as-caida20071105 &   0.0422 &    0.0049 &   0.0713 &    0.0051 &   0.7937 &    0.0051 \\
 p2p-Gnutella31   &   0.8146 &    0.0118 &   2.5231 &    0.0190 &   6.1104 &    0.0321 \\
 soc-Epinions1    &   0.8558 &    0.0208 &   2.9626 &    0.0214 &  12.7745 &    0.0203 \\
 web-Stanford     & 207.5102 &    0.0547 & 276.8251 &    0.0460 & 406.2662 &    0.0429 \\
 web-NotreDame    &   1.1191 &    0.0191 &  13.9595 &    0.0162 &  40.1689 &    0.0171 \\
 wiki-Talk        &   2.8892 &    0.1339 &   9.8011 &    0.1208 &  43.7422 &    0.1256 \\
 cit-Patents      & 337.1497 &    0.0907 & 375.2206 &    0.0590 & 396.8926 &    0.1143 \\
\bottomrule
\end{tabular}
\end{footnotesize}
\caption{Update times for 100 random edge deletions with $k \in \{1, 10, 100\}$ in directed complex networks. The columns ``static''  and ``dynamic'' contain the average time for the static and dynamic algorithm, respectively.}
\label{tbl:deletionComplexNetworkDirApp}
\end{table}

\begin{table}[H]
\centering
\begin{footnotesize}
\begin{tabular}{lrr|rr|rr}
\toprule
Graph & \multicolumn{2}{c}{k = 1} & \multicolumn{2}{c}{k = 10} & \multicolumn{2}{c}{k = 100}\\
           &    static [s] &   dynamic [s] &     static [s] &   dynamic [s] &     static [s] &   dynamic [s] \\
 \midrule \midrule
  HC-BIOGRID              &  0.0317 &    0.0036 &   0.0496 &    0.0044 &   0.1720 &    0.0071 \\
 Mus\_musculus            &  0.0086 &    0.0014 &   0.0261 &    0.0027 &   0.0989 &    0.0071 \\
 Caenorhabditis\_elegans  &  0.0101 &    0.0024 &   0.0209 &    0.0035 &   0.0950 &    0.0073 \\
 ca-GrQc                 &  0.0260 &    0.0015 &   0.0443 &    0.0020 &   0.1297 &    0.0033 \\
 advogato                &  0.0492 &    0.0044 &   0.1272 &    0.0062 &   0.3217 &    0.0090 \\
 hprd\_pp                 &  0.0339 &    0.0061 &   0.0692 &    0.0084 &   0.4139 &    0.0180 \\
 ca-HepTh                &  0.0656 &    0.0051 &   0.1278 &    0.0074 &   0.3346 &    0.0122 \\
 Drosophila\_melanogaster &  0.0410 &    0.0074 &   0.2310 &    0.0109 &   0.4046 &    0.0199 \\
 oregon1\_010526          &  0.0163 &    0.0042 &   0.0250 &    0.0056 &   0.2009 &    0.0135 \\
 oregon2\_010526          &  0.0235 &    0.0051 &   0.0445 &    0.0073 &   0.2601 &    0.0129 \\
 ca-HepPh                &  1.4666 &    0.0126 &   1.5403 &    0.0141 &   2.1020 &    0.0162 \\
 Homo\_sapiens            &  0.0497 &    0.0091 &   0.1217 &    0.0133 &   0.8283 &    0.0263 \\
 GoogleNw                &  0.0875 &    0.0096 &   2.3516 &    0.0157 &   3.3499 &    0.0203 \\
 CA-AstroPh              &  1.0323 &    0.0157 &   1.4989 &    0.0170 &   4.6072 &    0.0222 \\
 dip20090126\_MAX         &  1.6675 &    0.0187 &   2.4406 &    0.0301 &   4.5202 &    0.0439 \\
 CA-CondMat              &  0.1444 &    0.0107 &   0.3036 &    0.0142 &   1.1564 &    0.0245 \\
 email-Enron             &  0.3726 &    0.0217 &   0.6040 &    0.0201 &   1.9093 &    0.0300 \\
 loc-brightkite    &  0.2105 &    0.0352 &   2.3587 &    0.0586 &   4.3993 &    0.0964 \\
 gowalla           &  0.8376 &    0.1825 &   9.4024 &    0.2595 & 131.2249 &    0.3896 \\
 com-dblp        & 16.0662 &    0.4307 &  20.3018 &    0.5169 &  43.8383 &    0.8710 \\
 com-amazon      & 98.1340 &    0.6294 & 116.4843 &    1.1724 & 349.3862 &    1.6538 \\
 com-youtube     &  3.6342 &    1.0569 & 427.4117 &    3.6402 & 751.0728 &    4.7879 \\\bottomrule
\end{tabular}
\end{footnotesize}
\caption{Update times for 100 random edge removals with  $k \in \{1, 10, 100\}$ in undirected complex networks. The columns ``static''  and ``dynamic'' contain the average time for the static and dynamic algorithm, respectively.}
\label{tbl:deletionComplexNetworkUndirApp}
\end{table}

\FloatBarrier

\subsection{Speedups for Street Networks}
\label{app:street}
\begin{table}[H]
\centering
\begin{footnotesize}
\begin{tabular}{lrrr|rrr|rrr}
\toprule
Graph & \multicolumn{3}{c}{k = 1} & \multicolumn{3}{c}{k = 10} & \multicolumn{3}{c}{k = 100}\\
             &   gmean &    min &      max &   gmean &    min &      max &   gmean &   min &       max \\
\midrule \midrule
 maldives          &   485.9 &   9.3 &   734.5 &   482.7 &   3.7 &   883.5 &   584.8 &  10.9 &   980.4 \\
 faroe-islands     &   431.5 &   2.8 & 13250.3 &   276.5 &   1.7 & 11531.9 &   148.1 &   3.2 & 15783.5 \\
 liechtenstein     &   162.8 &   6.7 & 12125.3 &   159.5 &   2.2 & 15470.6 &    93.2 &   1.3 & 27578.6 \\
 isle-of-man       &   193.9 &   9.4 &  6855.9 &   117.3 &   1.2 &  7589.3 &    70.2 &   3.2 &   971.1 \\
 equatorial-guinea &   455.2 &   8.2 &  6440.1 &   453.2 &  37.1 & 16065.3 &   405.8 &  23.7 & 21588.5 \\
 malta             &   438.7 &  20.7 &  6884.2 &   519.3 &  11.0 & 14307.0 &   283.3 &   4.9 & 13639.7 \\
 belize            &   534.3 &  70.7 & 14497.9 &   502.0 &  19.2 & 25484.5 &   153.7 &   7.3 & 28820.4 \\
 azores            &  1181.4 &  86.6 &  4154.8 &  1257.0 &  63.2 &  6980.3 &  1167.9 &  33.3 &  5929.5 \\
 \midrule \midrule
 (geometric) mean  &   412.3 &  14.2 &  6191.9 &   372.5 &   7.3 &  9144.8 &   241.8 &   6.7 &  8220.6 \\
 \bottomrule
\end{tabular}
\end{footnotesize}
\caption{Speedups on recomputation over 100 edge insertions in directed street networks, for $k \in \{1, 10, 100\}$. The column ``gmean'' contains the geometric mean of the achieved speedups, ``min'' and ``max'' the minimum and the maximum speedup.}
\label{tbl:insertionsStreetNetworksDirected}
\end{table}

\begin{table}[H]
\centering
\begin{footnotesize}
\begin{tabular}{lrrr|rrr|rrr}
\toprule
Graph & \multicolumn{3}{c}{k = 1} & \multicolumn{3}{c}{k = 10} & \multicolumn{3}{c}{k = 100}\\
                &    gmean &   min &     max &    gmean &   min &     max &    gmean &   min &     max \\
\midrule \midrule
 maldives          &   285.3 &  26.0 &  481.6 &   311.2 &  17.1 &  564.5 &   308.0 &   7.6 &   663.7 \\
 faroe-islands     &   110.2 &   1.1 & 4115.4 &    61.6 &   1.7 & 4473.0 &    36.8 &   1.2 &  6374.2 \\
 liechtenstein     &    69.1 &   7.0 & 5274.2 &    37.9 &   4.6 & 8756.1 &    11.2 &   2.4 & 13198.3 \\
 isle-of-man       &    49.9 &   2.4 & 4410.2 &    36.2 &   4.0 & 4264.1 &    14.3 &   2.3 &  5294.5 \\
 equatorial-guinea &    86.8 &  16.8 & 4434.3 &    71.6 &   8.2 & 6111.2 &    55.1 &   4.1 &  7957.4 \\
 malta             &   107.1 &   3.9 & 2902.7 &   130.7 &   2.9 & 4338.3 &    62.3 &   2.9 &  8218.3 \\
 belize            &    86.2 &   5.3 & 5135.8 &    68.1 &  20.9 & 9255.8 &    23.6 &   4.0 & 10407.4 \\
 azores            &   220.9 &   9.1 & 1318.3 &   272.3 &   8.5 & 2563.5 &   218.0 &   8.7 &  1631.4 \\
 \midrule \midrule
  (geometric) mean  &   108.5 &   5.9 & 2821.6 &    90.7 &   6.2 & 3950.4 &    48.8 &   3.5 &  4892.4 \\
 \bottomrule
\end{tabular}
\end{footnotesize}
\caption{Speedups on recomputation over 100 edge insertions in undirected street networks, for $k \in \{1, 10, 100\}$. The column ``gmean'' contains the geometric mean of the achieved speedups, ``min'' and ``max'' the minimum and the maximum speedup.}
\label{tbl:insertionStreetNetworksUndirected}
\end{table}

\begin{table}[H]
\centering
\begin{footnotesize}
\begin{tabular}{lrrr|rrr|rrr}
\toprule
Graph & \multicolumn{3}{c}{k = 1} & \multicolumn{3}{c}{k = 10} & \multicolumn{3}{c}{k = 100}\\
             &   gmean &    min &      max &   gmean &    min &      max &   gmean &   min &       max \\
\midrule \midrule
 maldives          &   740.9 &  29.9 &  1227.1 &   795.1 &  25.8 &  1338.0 &   857.4 &  10.3 &  1655.0 \\
 faroe-islands     &   809.5 &   0.8 & 12583.8 &   697.2 &   0.7 & 13509.7 &   518.8 &   0.6 & 18936.6 \\
 liechtenstein     &   609.7 &  39.4 & 11082.1 &   382.0 &  27.4 & 24102.8 &   214.1 &   9.2 & 35979.0 \\
 isle-of-man       &   332.6 &  10.3 &   758.7 &   237.9 &   7.6 &   803.2 &   116.0 &   5.1 &   988.9 \\
 equatorial-guinea &  2018.3 & 165.4 & 14889.7 &  2333.4 &  78.0 & 28155.3 &  2098.5 &  25.1 & 35160.2 \\
 malta             &   688.6 & 169.2 &  8015.5 &   613.3 & 111.1 & 14959.3 &   392.5 &  32.7 & 17787.7 \\
 belize            &   769.2 &  11.8 & 14131.3 &   556.2 &   7.3 & 34348.3 &   292.8 &   4.2 & 46801.4 \\
 azores            &  2049.9 & 288.8 &  8574.0 &  2326.7 & 150.5 & 11610.8 &  1986.6 &  39.3 & 13917.1 \\
 \midrule \midrule
 (geometric) mean  &   847.7 &  31.1 &  6084.0 &   743.4 &  20.7 &  9357.4 &   519.3 &   9.0 & 12082.7 \\\bottomrule 
 \end{tabular}
\end{footnotesize}
\caption{Speedups on recomputation over 100 edge removals in directed street networks, for $k \in \{1, 10, 100\}$. The column ``gmean'' contains the geometric mean of the achieved speedups, ``min'' and ``max'' the minimum and the maximum speedup.}
\label{tbl:insertionsStreetNetworksDirected}
\end{table}

\begin{table}[H]
\centering
\begin{footnotesize}
\begin{tabular}{lrrr|rrr|rrr}
\toprule
Graph & \multicolumn{3}{c}{k = 1} & \multicolumn{3}{c}{k = 10} & \multicolumn{3}{c}{k = 100}\\
                &    gmean &   min &     max &    gmean &   min &     max &    gmean &   min &     max \\
\midrule \midrule
maldives          &   328.9 &  24.5 &  539.9 &   389.2 &  23.1 &  658.6 &   404.1 &   8.9 &   806.0 \\
 faroe-islands     &   123.6 &   0.2 & 1795.7 &    93.7 &   0.2 & 4421.8 &    55.5 &   0.3 &  6637.0 \\
 liechtenstein     &   170.5 &  16.3 & 4666.8 &    57.9 &  15.7 & 6926.5 &    13.8 &   6.0 & 11481.9 \\
 isle-of-man       &    94.2 &   2.4 & 4243.1 &    53.2 &   3.2 & 4610.9 &    11.7 &   2.4 &  5920.1 \\
 equatorial-guinea &   399.3 &  26.6 & 2919.2 &   406.2 &  28.4 & 4518.0 &   347.9 &   8.8 &  5519.6 \\
 malta             &   182.1 &  64.0 & 1306.5 &   144.3 &  54.9 & 1521.2 &    63.0 &  18.6 &  2144.7 \\
 belize            &    95.0 &   0.1 & 4535.5 &    54.0 &   0.1 & 6283.8 &    21.4 &   0.1 &  7672.6 \\
 azores            &   334.8 &  36.4 & 1314.7 &   326.8 &  28.8 & 1995.2 &   244.1 &  11.1 &  2319.9 \\
 \midrule \midrule
 (geometric) mean  &   187.2 &   5.2 & 2138.0 &   135.9 &   5.2 & 3076.0 &    67.2 &   2.9 &  4078.9 \\
 \bottomrule
  \end{tabular}
\end{footnotesize}
\caption{Speedups on recomputation over 100 edge removals in undirected street networks, for $k \in \{1, 10, 100\}$. The column ``gmean'' contains the geometric mean of the achieved speedups, ``min'' and ``max'' the minimum and the maximum speedup.}
\label{tbl:insertionStreetNetworksUndirected}
\end{table}

\FloatBarrier

\subsection{Running Times for Street Networks}
\label{app:time-street}

\begin{table}[H]
\centering
\begin{footnotesize}
\begin{tabular}{lrr|rr|rr}
\toprule
Graph & \multicolumn{2}{c}{k = 1} & \multicolumn{2}{c}{k = 10} & \multicolumn{2}{c}{k = 100}\\
                   &     static [s] &   dynamic [s] &     static [s] &   dynamic [s] &    static [s] &   dynamic [s] \\\midrule \midrule
  maldives          &  0.1959 &    0.0004 &  0.2038 &    0.0004 &  0.2897 &    0.0005 \\
 faroe-islands     &  4.0750 &    0.0094 &  4.2421 &    0.0153 &  4.0176 &    0.0271 \\
 liechtenstein     & 10.2787 &    0.0631 & 12.5527 &    0.0787 & 18.3544 &    0.1969 \\
 isle-of-man       &  6.4567 &    0.0333 &  7.0189 &    0.0598 &  8.3399 &    0.1188 \\
 equatorial-guinea & 15.7533 &    0.0346 & 15.8417 &    0.0350 & 19.8871 &    0.0490 \\
 malta             & 16.3053 &    0.0372 & 17.8396 &    0.0344 & 20.9484 &    0.0739 \\
 belize            & 32.5222 &    0.0609 & 34.8647 &    0.0695 & 47.5453 &    0.3092 \\
 azores            & 46.1730 &    0.0391 & 52.2209 &    0.0415 & 54.7890 &    0.0469 \\
  \bottomrule
\end{tabular}
\end{footnotesize}
\caption{Update times for 100 random edge insertions with  $k \in \{1, 10, 100\}$ in directed street networks. The columns ``static''  and ``dynamic'' contain the average time for the static and dynamic algorithm, respectively.}
\label{tbl:insertionComplexNetworkUndirApp}
\end{table}

\begin{table}[H]
\centering
\begin{footnotesize}
\begin{tabular}{lrr|rr|rr}
\toprule
Graph & \multicolumn{2}{c}{k = 1} & \multicolumn{2}{c}{k = 10} & \multicolumn{2}{c}{k = 100}\\
                   &     static [s] &   dynamic [s] &     static [s] &   dynamic [s] &    static [s] &   dynamic [s] \\\midrule \midrule
 maldives          &  0.1691 &    0.0006 &  0.2071 &    0.0007 &  0.2508 &    0.0008 \\
 faroe-islands     &  2.7337 &    0.0248 &  1.8672 &    0.0303 &  2.8906 &    0.0786 \\
 liechtenstein     &  9.5222 &    0.1377 & 11.2397 &    0.2963 & 21.6126 &    1.9293 \\
 isle-of-man       &  5.4850 &    0.1100 &  5.5143 &    0.1524 &  7.0543 &    0.4931 \\
 equatorial-guinea &  7.5368 &    0.0868 &  9.2345 &    0.1290 & 10.2358 &    0.1858 \\
 malta             &  9.0850 &    0.0849 & 10.0963 &    0.0772 & 13.6970 &    0.2198 \\
 belize            & 21.6908 &    0.2517 & 23.7571 &    0.3488 & 38.5866 &    1.6374 \\
 azores            & 17.4890 &    0.0792 & 19.3370 &    0.0710 & 25.3668 &    0.1163 \\
  \bottomrule
\end{tabular}
\end{footnotesize}
\caption{Update times for 100 random edge insertions with  $k \in \{1, 10, 100\}$ in undirected street networks. The columns ``static''  and ``dynamic'' contain the average time for the static and dynamic algorithm, respectively.}
\label{tbl:insertionComplexNetworkUndirApp}
\end{table}

\begin{table}[H]
\centering
\begin{footnotesize}
\begin{tabular}{lrr|rr|rr}
\toprule
Graph & \multicolumn{2}{c}{k = 1} & \multicolumn{2}{c}{k = 10} & \multicolumn{2}{c}{k = 100}\\
                   &     static [s] &   dynamic [s] &     static [s] &   dynamic [s] &    static [s] &   dynamic [s] \\\midrule \midrule
 maldives          &  0.2083 &    0.0003 &  0.2424 &    0.0003 &  0.2888 &    0.0003 \\
 faroe-islands     &  3.4412 &    0.0043 &  3.6236 &    0.0052 &  4.8986 &    0.0094 \\
 liechtenstein     & 11.2583 &    0.0185 & 12.0275 &    0.0315 & 17.9281 &    0.0838 \\
 isle-of-man       &  5.8541 &    0.0176 &  6.2707 &    0.0264 &  7.6060 &    0.0656 \\
 equatorial-guinea & 19.8018 &    0.0098 & 19.9973 &    0.0086 & 23.3713 &    0.0111 \\
 malta             & 15.1507 &    0.0220 & 14.7942 &    0.0241 & 18.3038 &    0.0466 \\
 belize            & 29.0321 &    0.0377 & 31.2502 &    0.0562 & 42.3123 &    0.1445 \\
 azores            & 43.8422 &    0.0214 & 45.3105 &    0.0195 & 51.5531 &    0.0260 \\
  \bottomrule
\end{tabular}
\end{footnotesize}
\caption{Update times for 100 random edge removals with  $k \in \{1, 10, 100\}$ in directed street networks. The columns ``static''  and ``dynamic'' contain the average time for the static and dynamic algorithm, respectively.}
\label{tbl:insertionComplexNetworkUndirApp}
\end{table}

\begin{table}[H]
\centering
\begin{footnotesize}
\begin{tabular}{lrr|rr|rr}
\toprule
Graph & \multicolumn{2}{c}{k = 1} & \multicolumn{2}{c}{k = 10} & \multicolumn{2}{c}{k = 100}\\
                   &     static [s] &   dynamic [s] &     static [s] &   dynamic [s] &    static [s] &   dynamic [s] \\\midrule \midrule
 maldives          &  0.1689 &    0.0005 &  0.2073 &    0.0005 &  0.2535 &    0.0006 \\
 faroe-islands     &  1.8495 &    0.0150 &  2.0199 &    0.0216 &  3.0493 &    0.0549 \\
 liechtenstein     &  9.2484 &    0.0542 & 10.2024 &    0.1761 & 17.7007 &    1.2800 \\
 isle-of-man       &  4.7809 &    0.0508 &  5.1547 &    0.0968 &  6.6677 &    0.5684 \\
 equatorial-guinea &  9.2114 &    0.0231 & 10.0027 &    0.0246 & 12.8164 &    0.0368 \\
 malta             &  8.6663 &    0.0476 &  9.1300 &    0.0633 & 12.4407 &    0.1975 \\
 belize            & 16.8559 &    0.1775 & 18.9041 &    0.3498 & 30.4396 &    1.4236 \\
 azores            & 16.5104 &    0.0493 & 17.9961 &    0.0551 & 23.8098 &    0.0975 \\
  \bottomrule
\end{tabular}
\end{footnotesize}
\caption{Update times for 100 random edge removals with  $k \in \{1, 10, 100\}$ in undirected street networks. The columns ``static''  and ``dynamic'' contain the average time for the static and dynamic algorithm, respectively.}
\label{tbl:insertionComplexNetworkUndirApp}
\end{table}

\end{document}